\documentclass[prc,preprint,showpacs,showkeys,lengthcheck,
  nofootinbib,tightenlines,onecolumn,notitlepage,preprintnumbers,superscriptaddress]{revtex4-1}

\usepackage[utf8]{inputenc}
\usepackage{newtxtext,newtxmath}
\usepackage[mathcal]{euscript}

\usepackage{graphicx}
\usepackage[usenames,dvipsnames]{xcolor}
\graphicspath{{figs/}}

\usepackage{hyperref}
\hypersetup{colorlinks=true,citecolor=blue,linkcolor=blue,urlcolor=blue}

\usepackage{diagbox}

\usepackage{amsmath,amssymb,amsfonts}
\usepackage{slashed}
\usepackage{bm}

\newcommand{\tddr}[1]{\textcolor{red}{#1}} 

\begin{document}

\title{Polynomiality sum rules for generalized parton distributions of spin-1 targets}

\author{W. Cosyn}
\affiliation{Department of Physics and Astronomy, Ghent University, 
             Proeftuinstraat 86, B9000 Ghent, Belgium}
\email{wim.cosyn@ugent.be}
\author{A. Freese}
\affiliation{Argonne National Laboratory, Lemont, Illinois 60439, USA}
\email{afreese@anl.gov}
\author{B. Pire}
\affiliation{Centre de Physique Th\'{e}orique, CNRS, \'{E}cole Polytechnique, I. P. Paris, F-91128 Palaiseau, France}
\email{bernard.pire@polytechnique.edu}

\begin{abstract}
  We present the polynomiality sum rules for all leading-twist quark and gluon generalized parton distributions (GPDs) of spin-1 targets such as the deuteron nucleus.  The sum rules connect the Mellin moments of these GPDs to polynomials in skewness parameter $\xi$, which contain generalized form factors as their coefficients.  The decompositions of local currents in terms of generalized form factors for spin-1 targets are obtained as a by-product of this derivation.  
\end{abstract}

\maketitle

\onecolumngrid



\section{Introduction}
\label{sec:intro}

Generalized parton distributions encode nonperturbative QCD dynamics of hadrons and appear in the Lorentz-covariant decomposition of off-forward matrix elements of quark and gluon correlators~\cite{Diehl:2003ny,Belitsky:2005qn}.  Due to QCD factorization theorems, these generalized parton distributions (GPDs) appear as the soft part in scattering amplitudes of hard exclusive processes such as deeply virtual Compton scattering and deeply virtual meson production.
The GPDs permit a physical interpretation as a partonic density
and more specifically can be related by a 2D Fourier transform to three-dimensional
(light cone momentum + transverse spatial coordinates)
partonic distributions~\cite{Burkardt:2000za,Ralston:2001xs,Burkardt:2002hr,Diehl:2002he,Diehl:2005jf}.

Lorentz covariance yields polynomiality sum rules for the Mellin moments of GPDs, where the $s$th Mellin moment corresponds to an integral over $x^{s-1}$ times the GPD.
This remarkable property arises in the following way.  Mellin moments of the bilocal light cone operators appearing in the parton correlators lead to towers of local operators.  Off-forward matrix elements of these local operators are parametrized in terms of generalized form factors (GFFs).  The $s$th Mellin moments of GPDs then correspond 
to finite polynomials
in the skewness variable $\xi$ [see Eq.~(\ref{eq:xi})], with the GFFs appearing linearly in the coefficients.  These polynomiality conditions impose strong constraints on the modeling of the corresponding GPDs.
    
First Mellin moments of GPDs result in regular form factors of local currents.  Second Mellin moments of helicity conserving GPDs can be connected to the gravitational form factors, which appear in the Lorentz-covariant decomposition of the energy-momentum tensor and can be used to study the spin, mass and pressure properties of hadrons~\cite{Ji:1996nm,Polyakov:2002yz,Taneja:2011sy,Polyakov:2018zvc,Burkert:2018bqq,Lorce:2018egm}.
    
As GFFs appear in the decomposition of local operators, they can be calculated on the lattice~\cite{Hagler:2003jd,Gockeler:2003jfa,Gockeler:2006zu,Brommel:2007xd,Winter:2017bfs,Detmold:2017oqb,*PhysRevD.99.019904}.  From the computation of these local matrix elements, after performing an inverse Mellin transform, the GPDs or collinear parton distribution functions (in the case of forward matrix elements) can in principle be recovered.
    
The GFF decomposition for spin-1/2 targets and the resulting polynomiality sum rules for the leading-twist GPDs have been extensively covered in the literature~\cite{Ji:1998pc,Ji:2000id,Hagler:2004yt,Chen:2004cg}.
For spins 0 and 1, there has not been a similar systematic study,
but for spin-0 the decompositions of local currents and polynomiality sum rules
are comparatively simple and results can be found in literature on pion GFFs
(see for instance~\cite{Broniowski:2008dy,Dorokhov:2011ew}).
For spin-1 the picture is still incomplete.
While the leading twist helicity conserving GPDs for a spin-1 target were introduced and studied quite some time ago~\cite{Berger:2001zb,Cano:2003ju}, the twist-2 transversity GPDs for spin 1 were only introduced recently~\cite{Cosyn:2018rdm}. In this article we present a systematic derivation of the polynomiality sum rules for all leading-twist quark and gluon GPDs of spin-1 targets with the GFF decomposition of local operators for spin-1 targets as an important byproduct of this derivation.  

A complete picture of polynomiality for spin-1 targets is desirable due to
an emerging interest in the partonic structure of particular spin-1 targets,
and hard exclusive reactions involving such targets.
Generalized parton distributions of the
deuteron~\cite{Berger:2001zb,Cano:2002ph,Kirchner:2002zu,Cano:2003ju,Kirchner:2003wt,Amrath:2007zz, Anikin:2010ng,Anikin:2011aa,Dong:2014eya,Cosyn:2018rdm},
rho meson~\cite{Sun:2017gtz,Sun:2018tmk,Adhikari:2018umb,Sun:2018ldr},
phi meson~\cite{Detmold:2017oqb},
and photon~\cite{Friot:2006mm,Gabdrakhmanov:2012aa,Mukherjee:2013yf}
have been the subjects of theoretical studies.
Deeply virtual Compton scattering from the deuteron has been
performed at HERMES~\cite{Movsisyan:2010pia,Airapetian:2010aa},
and is a topic of interest at Jefferson Lab~\cite{Mazouz:2007aa,Armstrong:2017wfw},
as well as the proposed Electron Ion Collider~\cite{Accardi:2012qut}.
Generalized distribution amplitudes (GDAs, see Refs.~\cite{Diehl:1998dk,Diehl:2000uv} for details) for the rho-rho meson pair~\cite{Anikin:2003fr,Anikin:2005ur} (which
can be probed by the crossed reaction $\gamma^*\gamma\rightarrow \rho \rho$) could
potentially be studied at Belle II.
Peculiar aspects of diphoton GDAs have also been discussed~\cite{ElBeiyad:2008ss}.

The material in this article is organized as follows.  We restate the leading-twist quark and gluon light cone correlator decompositions in GPDs for spin-1 targets in Sec.~\ref{sec:gpd}. In Sec.~\ref{sec:mellin}, the general relation between the Mellin moments of bilocal gauge-invariant light cone operators and local operators is rederived.  In Sec.~\ref{sec:gff}, based on the method developed by Ji and Lebed~\cite{Ji:2000id}, we discuss the counting of the number of GFFs that appear in the decomposition of the local operators found in Sec.~\ref{sec:mellin} based on symmetries and selection rules.  This counting provides an important check on the further results in this paper.  Section~\ref{sec:polynomial} contains the main results of this paper, viz., the decomposition of the local operators in GFFs and the resulting polynomiality sum rules for the GPDs of spin-1 targets.  We find agreement between our decompositions and the counting established in Sec.~\ref{sec:gff}.  Finally, conclusions are stated in Sec.~\ref{sec:concl}.  Appendix~\ref{sec:conventions} summarizes the conventions we use in this work, and Appendix~\ref{sec:comp} outlines the correspondence between this work and a second gluon spin-1 GFF decomposition in the literature~\cite{Detmold:2017oqb,*PhysRevD.99.019904}. The connection with the gravitational form factors of spin-1 targets is left for a future study~\cite{Cosyn:2019aio}.

\section{Leading-twist GPDs of spin-1 hadrons}
\label{sec:gpd}
Quark and gluon GPDs are defined as Lorentz-invariant functions appearing
in the decompositions of light cone correlators\cite{Diehl:2003ny,Belitsky:2005qn}.
GPDs are classified by their collinear twist,
which is equal to their dimension minus the projection
of their spin onto the lightlike vector $n$ defining the
``plus'' direction~\cite{Braun:2003rp}.
GPDs of higher twist\footnote{Higher-twist GPDs are needed to ensure the QED gauge invariance of the amplitude \cite{Anikin:2010ng,Anikin:2011aa}.} make smaller contributions to cross sections
for hard processes such as deeply virtual Compton scattering,
and at high $Q^2$ the lowest-twist GPDs dominate the cross section.
The lowest-twist GPDs are twist-2 and are often called leading twist.
The leading-twist GPDs are specifically what we consider here.

There are three leading-twist quark correlators,
which are defined by the following off-forward matrix elements (conventions used in this work are summarized in Appendix~\ref{sec:conventions}):
\begin{subequations}
  \begin{align}
    V^q_{\lambda^\prime\lambda}
    &=
    \int_{-\infty}^\infty \frac{\mathrm{d}\kappa}{2\pi}
    e^{2ix(P  n)\kappa}
    \langle p^\prime, \lambda^\prime |
    \bar{q}(-n\kappa)
    \slashed{n}
    [-n\kappa,n\kappa]
    q(n\kappa)
    | p, \lambda \rangle \,,
    \\
    A^q_{\lambda^\prime\lambda}
    &=
    \int_{-\infty}^\infty \frac{\mathrm{d}\kappa}{2\pi}
    e^{2ix(P  n)\kappa}
    \langle p^\prime, \lambda^\prime |
    \bar{q}(-n\kappa)
    \slashed{n} \gamma_5
    [-n\kappa,n\kappa]
    q(n\kappa)
    | p, \lambda \rangle \,,
    \\
    T^q_{\lambda^\prime\lambda}
    &=
    \int_{-\infty}^\infty \frac{\mathrm{d}\kappa}{2\pi}
    e^{2ix(P  n)\kappa}
    \langle p^\prime, \lambda^\prime |
    \bar{q}(-n\kappa)
    \sigma^{ni}
    [-n\kappa,n\kappa]
    q(n\kappa)
    | p, \lambda \rangle \,,
  \end{align}
  \label{eqn:gpd:quark}
\end{subequations}
where
\begin{align}
  \left[x,y\right]
  =
  \mathcal{P}
  \exp\left\{
    ig\int_{y}^{x} A(z)  \,\mathrm{d}z
    \right\}
\end{align}
is a Wilson line from $x$ to $y$, with $\mathcal{P}$ signifying path ordering.
There are also three leading-twist gluon correlators, which are defined by
\begin{subequations}
  \begin{align}
    V^g_{\lambda^\prime\lambda}
    &=
    \frac{2}{(P  n)}
    \int_{-\infty}^\infty \frac{\mathrm{d}\kappa}{2\pi}
    e^{2ix(P  n)\kappa}
    \langle p^\prime, \lambda^\prime |
   2 \mathrm{Tr}\left\{
      [n\kappa,-n\kappa]
      G^{n\pi}(-n\kappa)
      [-n\kappa,n\kappa]
      G_{\pi n}(n\kappa)
      \right\}
    | p, \lambda \rangle \,,
    \\
    A^g_{\lambda^\prime\lambda}
    &=
    -i\frac{2}{(P  n)}
    \int_{-\infty}^\infty \frac{\mathrm{d}\kappa}{2\pi}
    e^{2ix(P  n)\kappa}
    \langle p^\prime, \lambda^\prime |
   2 \mathrm{Tr}\left\{
      [n\kappa,-n\kappa]
      G^{n\pi}(-n\kappa)
      [-n\kappa,n\kappa]
      \widetilde{G}_{\pi n}(n\kappa)
      \right\}
    | p, \lambda \rangle \,,
    \\
    T^g_{\lambda^\prime\lambda}
    &=
    \underset{\{ij\}}{\mathcal{S}}
    \frac{2}{(P  n)}
    \int_{-\infty}^\infty \frac{\mathrm{d}\kappa}{2\pi}
    e^{2ix(P  n)\kappa}
    \langle p^\prime, \lambda^\prime |
   2 \mathrm{Tr}\left\{
      [n\kappa,-n\kappa]
      G^{ni}(-n\kappa)
      [-n\kappa,n\kappa]
      G^{jn}(n\kappa)
      \right\}
    | p, \lambda \rangle \,,
  \end{align}
  \label{eqn:gpd:gluon}
\end{subequations}
where the symmetrization operator $\mathcal{S}$ is defined in Eq.~(\ref{eq:symm}).
Here, and elsewhere in this work, the indices $i$ and $j$
signify transverse light cone coordinates.
Each of these correlators has an additional dependence on the
renormalization scale $\mu^2$, which we have not notated for brevity.

The correlators defined in Eqs.~(\ref{eqn:gpd:quark}) and (\ref{eqn:gpd:gluon})
apply to any hadron, but the number of independent Lorentz structures this
can be decomposed into, and thus the number of GPDs a hadron has,
depends on the hadron's spin.
Here, we give the decompositions of the light cone correlators for spin-1
specifically.
The vector quark and gluon correlators have the following decomposition:
\begin{multline}
  V_{\lambda^\prime\lambda}^i
  =
  -(\epsilon'^{*} \epsilon) H_1^i(x,\xi,t)
  + \frac{
    (\epsilon  n)(\epsilon'^{*}  P)+(\epsilon'^{*}  n)(\epsilon  P)
  }{(P  n)} H_2^i(x,\xi,t)
  - \frac{2(\epsilon  P)(\epsilon'^{*}  P)}{M^2} H_3^i(x,\xi,t)
  \\
  + \frac{
    (\epsilon  n)(\epsilon'^{*}  P)-(\epsilon'^{*}  n)(\epsilon  P)
  }{(P  n)} H_4^i(x,\xi,t)
  + \left[ \frac{M^2(\epsilon  n)(\epsilon'^{*}  n)}{(P  n)^2}
    + \frac{1}{3}(\epsilon \epsilon'^{*})\right] H_5^i(x,\xi,t) \qquad \qquad \text{for} \quad i=q,g\,,
  \label{eqn:gpd:vector}
\end{multline}
while the axial vector quark and gluon correlators decompose as
\begin{multline}
  A_{\lambda^\prime\lambda}^i
  =
  - \frac{i\epsilon_{n\epsilon'^{*}\epsilon P}}{(P  n)} \widetilde{H}_1^i(x,\xi,t)
  + \frac{2i\epsilon_{n\Delta P \pi}}{M^2}
  \frac{\epsilon^\pi(\epsilon'^{*}  P) + \epsilon'^{*\pi}(\epsilon  P)}{(P  n)}
  \widetilde{H}_2^i(x,\xi,t)
  \\
  + \frac{2i\epsilon_{n\Delta P \pi}}{M^2}
  \frac{\epsilon^\pi(\epsilon'^{*}  P) - \epsilon'^{*\pi}(\epsilon  P)}{(P  n)}
  \widetilde{H}_3^i(x,\xi,t)
  + \frac{i\epsilon_{n\Delta P \pi}}{2(P  n)}
  \frac{\epsilon^\pi(\epsilon'^{*}  n) + \epsilon'^{*\pi}(\epsilon  n)}{(P  n)}
  \widetilde{H}_4^i(x,\xi,t)
  \qquad \qquad \text{for} \quad i=q,g\,,
  \label{eqn:gpd:axial}
\end{multline}
where we use the shorthand
$\epsilon_{xyzw} = \epsilon_{\mu\nu\rho\sigma} x^\mu y^\nu z^\rho w^\sigma$
whenever $x,y,z,w$ are four-vectors.
These were first found in Ref.~\cite{Berger:2001zb}, though we define $P$
to be half of the $P$ used in the same source, producing some superficial
differences in the formulas.

The decompositions for the quark and gluon transversities for spin-1 hadrons
were found later, in Ref.~\cite{Cosyn:2018rdm}.
For the quark transversity, we have
\begin{align}
& T_{\lambda^\prime\lambda}^q =
M\frac{(\epsilon'^{*}n)\epsilon^i-\epsilon'^{*i}(\epsilon n)} {2\sqrt{2}(Pn)}
H^{qT}_1(x,\xi,t)\nonumber\\
&\qquad+M\left[\frac{2P^i(\epsilon n)(\epsilon'^{*} 
n)}{2\sqrt{2}(Pn)^2}-\frac{(\epsilon 
n)\epsilon'^{i*}+\epsilon^i(\epsilon'^{*} n)}{2\sqrt{2}(Pn)}\right] 
H^{qT}_2(x,\xi,t)\nonumber\\
&\qquad+\left[ \frac{(\epsilon'^{*} n)\Delta^i-\epsilon'^{i*}(\Delta n)}{M(Pn)}(\epsilon 
P) - \frac{(\epsilon n)\Delta^i-\epsilon^i (\Delta n)}{M(Pn)}(\epsilon'^{*}P)\right]  H^{qT}_3(x,\xi,t)\nonumber\\
&\qquad+\left[ \frac{(\epsilon'^{*} n)\Delta^i-\epsilon'^{i*}(\Delta n)}{M(Pn)}(\epsilon 
P) + \frac{(\epsilon n)\Delta^i-\epsilon^i (\Delta n)}{M(Pn)}(\epsilon'^{*}P)\right]  H^{qT}_4(x,\xi,t)\nonumber\\
&\qquad+ M\left[ \frac{(\epsilon'^{*} n)\Delta^i-\epsilon'^{i*}(\Delta n)}{2\sqrt{2}(Pn)^2}(\epsilon 
n) + \frac{(\epsilon n)\Delta^i-\epsilon^i (\Delta n)}{2\sqrt{2}(Pn)^2}(\epsilon'^{*}n)\right] 
H^{qT}_5(x,\xi,t)
\nonumber\\
&\qquad+ \frac{(\Delta^i+2\xi P^i)}{M}(\epsilon'^{*} \epsilon)H^{qT}_6(x,\xi,t)
+\frac{(\Delta^i+2\xi P^i)}{M} \frac{(\epsilon'^{*} P)(\epsilon P)}{M^2}
H^{qT}_7(x,\xi,t)\nonumber\\
&\qquad + \left[ \frac{(\epsilon'^{*} n)P^i-\epsilon'^{i*}(Pn)}{M(Pn)}(\epsilon 
P) + \frac{(\epsilon n)P^i-\epsilon^i (Pn)}{M(Pn)}(\epsilon'^{*}P)\right] 
H^{qT}_8(x,\xi,t)\nonumber\\
&\qquad + \left[ \frac{(\epsilon'^{*} n)P^i-\epsilon'^{i*}(Pn)}{M(Pn)}(\epsilon 
P) - \frac{(\epsilon n)P^i-\epsilon^i (Pn)}{M(Pn)}(\epsilon'^{*}P)\right] 
H^{qT}_9(x,\xi,t)\,,
  \label{eqn:gpd:transversity:quark}
\end{align}
and for the gluon transversity
\begin{align}
& T_{\lambda^\prime\lambda}^g =
\underset{\{ij\}}{\mathcal{S}} \left\{
(\Delta^i+2\xi P^i)\frac{(\epsilon'^{*}n)\epsilon^j-\epsilon'^{j*}(\epsilon 
n)} {2\sqrt{2}(Pn)}
H^{gT}_1(x,\xi,t)\right.\nonumber\\
&\qquad+(\Delta^i+2\xi P^i)\left[\frac{2P^j(\epsilon n)(\epsilon'^{*} 
n)}{2\sqrt{2}(Pn)^2}-\frac{(\epsilon 
n)\epsilon'^{j*}+\epsilon^j(\epsilon'^{*} 
n)}{2\sqrt{2}(Pn)}\right]H^{gT}_2(x,\xi,t)\nonumber\\
&\qquad+\frac{(\Delta^i+2\xi P^i)}{M}\left[ \frac{(\epsilon'^{*} n)\Delta^j-\epsilon'^{j*}(\Delta n)}{M(Pn)}(\epsilon 
P) - \frac{(\epsilon n)\Delta^j-\epsilon^j (\Delta n)}{M(Pn)}(\epsilon'^{*}P)\right] H^{gT}_3(x,\xi,t)\nonumber\\
&\qquad+\frac{(\Delta^i+2\xi P^i)}{M}\left[ \frac{(\epsilon'^{*} n)\Delta^j-\epsilon'^{j*}(\Delta n)}{M(Pn)}(\epsilon 
P) + \frac{(\epsilon n)\Delta^j-\epsilon^j (\Delta n)}{M(Pn)}(\epsilon'^{*}P)\right]  H^{gT}_4(x,\xi,t)\nonumber\\
&\qquad -\left[\frac{(\epsilon'^{*} 
n)P^i-(Pn)\epsilon'^{i*}}{(Pn)}\right]\left[\frac{(\epsilon 
n)P^j-(Pn)\epsilon^j}{(Pn)}\right]H^{gT}_5(x,\xi , t)
\nonumber\\
&\qquad+  \left[\frac{(\epsilon'^{*} 
n)\Delta^i-(\Delta n)\epsilon'^{i*}}{2(Pn)}\right]\left[\frac{(\epsilon 
n)\Delta^j-(\Delta n)\epsilon^j}{2(Pn)}\right]H^{gT}_6(x,\xi,t)
+\frac{(\Delta^i+2\xi P^i)}{M}\frac{(\Delta^j+2\xi P^j)}{M} 
\frac{(\epsilon'^{*} 
P)(\epsilon P)}{M^2}
H^{gT}_7(x,\xi,t)\nonumber\\
&\qquad + \frac{\Delta^i+2\xi P^i}{M}\left[ \frac{(\epsilon'^{*} 
n)P^j-\epsilon'^{j*}(Pn)}{M(Pn)}(\epsilon 
P) + \frac{(\epsilon n)P^j-\epsilon^j (Pn)}{M(Pn)}(\epsilon'^{*}P)\right] 
H^{gT}_8(x,\xi,t)\nonumber\\
&\qquad \left. + \frac{\Delta^i+2\xi P^i}{M} \left[ \frac{(\epsilon'^{*} 
n)P^j-\epsilon'^{j*}(Pn)}{M(Pn)}(\epsilon 
P) - \frac{(\epsilon n)P^j-\epsilon^j (Pn)}{M(Pn)}(\epsilon'^{*}P)\right] 
H^{gT}_9(x,\xi,t)\right\}\,.
  \label{eqn:gpd:transversity:gluon}
\end{align}

\section{Mellin moments of bilocal operators}
\label{sec:mellin}
In this section, we consider the relationship between Mellin moments
of generalized parton distributions and matrix elements of local currents.
The Mellin moments of the GPDs defined in Sec.~\ref{sec:gpd}
can be found by evaluating the Mellin moments of the bilocal operators
that define the correlators in Eqs.~(\ref{eqn:gpd:quark}) and (\ref{eqn:gpd:gluon}).
The three quark operators of interest here are
\begin{subequations}
  \begin{align}
    \mathcal{O}_{qV}
    &=
    \int_{-\infty}^\infty \frac{\mathrm{d}\kappa}{2\pi}
    e^{2ix(P  n)\kappa}
    \bar{q}(-n\kappa)
    \slashed{n}
    [-n\kappa,n\kappa]
    q(n\kappa) \,,
    \\
    \mathcal{O}_{qA}
    &=
    \int_{-\infty}^\infty \frac{\mathrm{d}\kappa}{2\pi}
    e^{2ix(P  n)\kappa}
    \bar{q}(-n\kappa)
    \slashed{n} \gamma_5
    [-n\kappa,n\kappa]
    q(n\kappa) \,,
    \\
    \mathcal{O}_{qT}
    &=
    \int_{-\infty}^\infty \frac{\mathrm{d}\kappa}{2\pi}
    e^{2ix(P  n)\kappa}
    \bar{q}(-n\kappa)
    \sigma^{ni}
    [-n\kappa,n\kappa]
    q(n\kappa) \,,
  \end{align}
\end{subequations}
while the gluon operators are
\begin{subequations}
  \begin{align}
    \mathcal{O}_{gV}
    &=
    \frac{2}{(P  n)}
    \int_{-\infty}^\infty \frac{\mathrm{d}\kappa}{2\pi}
    e^{2ix(P  n)\kappa}
    2 \mathrm{Tr}\left\{
      [n\kappa,-n\kappa]
      G^{n\pi}(-n\kappa)
      [-n\kappa,n\kappa]
      G_{\pi n}(n\kappa)
      \right\} \,,
    \\
    \mathcal{O}_{gA}
    &=
    -i\frac{2}{(P  n)}
    \int_{-\infty}^\infty \frac{\mathrm{d}\kappa}{2\pi}
    e^{2ix(P  n)\kappa}
    2 \mathrm{Tr}\left\{
      [n\kappa,-n\kappa]
      G^{n\pi}(-n\kappa)
      [-n\kappa,n\kappa]
      \widetilde{G}_{\pi n}(n\kappa)
      \right\} \,,
    \\
    \mathcal{O}_{gT}
    &=
    \underset{\{ij\}}{\mathcal{S}}
    \frac{2}{(P  n)}
    \int_{-\infty}^\infty \frac{\mathrm{d}\kappa}{2\pi}
    e^{2ix(P  n)\kappa}
    2 \mathrm{Tr}\left\{
      [n\kappa,-n\kappa]
      G^{ni}(-n\kappa)
      [-n\kappa,n\kappa]
      G^{jn}(n\kappa)
      \right\} \,.
  \end{align}
\end{subequations}
Like the correlators they produce through their off-forward matrix elements,
these operators have an implicit dependence on a renormalization scale $\mu^2$.

The identity $x^s\int \mathrm{d}\kappa\, e^{2ix(P  n)\kappa} f(\kappa)
= \left(\frac{i}{2(P  n)}\right)^s \int \mathrm{d}\kappa\, e^{2ix(P  n)\kappa}f^{(s)}(\kappa)$
can be used in the Mellin transforms of the bilocal operators
and combined with the Leibniz product rule and chain rule
to recast $x^s$ in terms of the gauge-covariant derivative.
The action of the covariant derivative depends on which representation
of $\mathrm{SU}(3,\mathbb{C})$ the object being differentiated transforms under,
and its actions on the field operators of interest are
\begin{subequations}
  \begin{align}
    \mathcal{D}_\mu q(x)
    &= \partial_\mu q(x) - ig A_\mu(x) q(x) \,,
    \\
    \mathcal{D}_\mu \bar{q}(x)
    &= \partial_\mu \bar{q}(x) + ig \bar{q}(x) A_\mu(x) \,,
    \\
    \mathcal{D}_\mu G_{\nu\pi}(x)
    &= \partial_\mu G_{\nu\pi}(x) - ig [A_\mu(x),G_{\nu\pi}(x)] \,.
  \end{align}
\end{subequations}
If we define a two-sided covariant derivative using
\begin{align}
  g(y) \overleftrightarrow{\mathcal{D}}_\mu f(x)
  & \equiv
  \frac{1}{2}\Big(
  g(y) [y,x] \left( \mathcal{D}_\mu f(x) \right)
  -
  \left( \mathcal{D}_\mu g(y) \right) [y,x] f(x)
  \Big)
\end{align}
(where we have absorbed the gauge link in order to make the notation
less cumbersome),
then we find that
\begin{subequations}
  \begin{align}
    \frac{\mathrm{d}^s}{\mathrm{d}\kappa^s}\left[
      \bar{q}(-n\kappa)
      \Gamma
      [-n\kappa,n\kappa]
      q(n\kappa)
      \right]
    &=
    2^s
    \bar{q}(-n\kappa)
    \Gamma
    (n \overleftrightarrow{\mathcal{D}})^s
    q(n\kappa) \,,
    \\
    \frac{\mathrm{d}^s}{\mathrm{d}\kappa^s}\left[
      \mathrm{Tr}\left\{
        [n\kappa,-n\kappa]
        \mathcal{G}(-n\kappa)
        [-n\kappa,n\kappa]
        \mathcal{G}^\prime(n\kappa)
        \right\}
      \right]
    &=
    2^s
    \mathrm{Tr}\left\{
      [n\kappa,-n\kappa]
      \mathcal{G}(-n\kappa)
      (n \overleftrightarrow{\mathcal{D}})^s
      \mathcal{G}^\prime(n\kappa)
      \right\} \,,
  \end{align}
\end{subequations}
where $\Gamma$ is a generic Clifford matrix, and we have used
$\mathcal{G}$ and $\mathcal{G}^\prime$ to denote generic components
of the gluon field strength or its dual.
Lastly, if we define an auxiliary variable $\lambda = 2(P  n)x$,
then one has
\begin{align}
  \int \mathrm{d}x\, e^{2ix(P  n)\kappa}
  =
  \frac{1}{2(P  n)}\int \mathrm{d}\lambda e^{i\lambda\kappa}
  =
  \frac{2\pi}{2(P  n)} \delta(\kappa)
  ,
\end{align}
and the Mellin moments of generic quark and gluon operators become
\begin{subequations}
  \begin{align}
    \int_{-1}^1
    x^s \mathcal{O}_{qX}
    \,\mathrm{d}x
    &=
    \frac{1}{2}
    \frac{
      n_{\mu_1} \ldots n_{\mu_s}
    }{(P  n)^{s+1}}
    \bar{q}(0)
    \Gamma
    \left(i\overleftrightarrow{\mathcal{D}}^{\mu_1}\right)
    \ldots
    \left(i\overleftrightarrow{\mathcal{D}}^{\mu_s}\right)
    q(0)
    \equiv
    \frac{
      n_{\mu} n_{\mu_1} \ldots n_{\mu_s}
    }{(P  n)^{s+1}}
    \mathcal{O}_{qX}^{\mu\mu_1\ldots\mu_s} \,,
    \\
    \int_{-1}^1 x^s \mathcal{O}_{gX}
    \,\mathrm{d}x
    &=
    \frac{
      n_{\mu_1} \ldots n_{\mu_s}
    }{(P  n)^{s+2}}
    2\mathrm{Tr}\left\{
      \mathcal{G}(0)
      \left(i\overleftrightarrow{\mathcal{D}}^{\mu_1}\right)
      \ldots
      \left(i\overleftrightarrow{\mathcal{D}}^{\mu_s}\right)
      \mathcal{G}^\prime(0)
      \right\}
    \equiv
    \frac{
      n_{\mu} n_{\nu} n_{\mu_1} \ldots n_{\mu_s}
    }{(P  n)^{s+2}}
    \mathcal{O}_{gX}^{\mu\nu\mu_1\ldots\mu_s}
    \,,
  \end{align}
  \label{eqn:local}
\end{subequations}
where additional instances of $n$ are pulled out of the operator,
having come from the definitions of the light cone correlators.
The Mellin moments of the bilocal operator have thus
become local operators.

It is worth noting that, since $n_{\mu_1}\ldots n_{\mu_s}$ is symmetric
and traceless, we can additionally symmetrize and subtract the trace
of the matrices between the quark or gluon fields
in Eqs.~(\ref{eqn:local})---that is, we can apply the
operator $\mathcal{S}$ [see Eq.~(\ref{eq:symm})]---without changing the result.
Moreover, since the actual matrices (quark case) or field operators
(gluon case) are contracted with $n$ at leading twist,
the results for specific correlators can be symmetrized further.
In particular, for quarks we find
\begin{subequations}
  \begin{align}
    \mathcal{O}_{qV}^{\mu\mu_1\ldots\mu_s}
    &=
    \frac{1}{2}
    \underset{\{\mu\mu_1\ldots\mu_s\}}{\mathcal{S}}
    \bar{q}(0)
    \gamma^{\mu}
    \left(i\overleftrightarrow{\mathcal{D}}^{\mu_1}\right)
    \ldots
    \left(i\overleftrightarrow{\mathcal{D}}^{\mu_s}\right)
    q(0) \,,
    \label{eqn:local:quark:unpolarized}
    \\
    \mathcal{O}_{qA}^{\mu\mu_1\ldots\mu_s}
    &=
    \frac{1}{2}
    \underset{\{\mu\mu_1\ldots\mu_s\}}{\mathcal{S}}
    \bar{q}(0)
    \gamma^{\mu}
    \gamma_5
    \left(i\overleftrightarrow{\mathcal{D}}^{\mu_1}\right)
    \ldots
    \left(i\overleftrightarrow{\mathcal{D}}^{\mu_s}\right)
    q(0) \,,
    \label{eqn:local:quark:polarized}
    \\
    \mathcal{O}_{qT}^{\mu\nu\mu_1\ldots\mu_s}
    &=
    \frac{1}{2}
    \underset{[\mu\nu]}{\mathcal{A}}
    \underset{\{\mu\mu_1\ldots\mu_s\}}{\mathcal{S}}
    \bar{q}(0)
    \sigma^{\mu\nu}
    \left(i\overleftrightarrow{\mathcal{D}}^{\mu_1}\right)
    \ldots
    \left(i\overleftrightarrow{\mathcal{D}}^{\mu_s}\right)
    q(0) \,.
    \label{eqn:local:quark:transversity}
  \end{align}
  \label{eqn:local:quark}
\end{subequations}
Here, $\mathcal{A}$ is defined in Eq.~(\ref{eq:antisymm}) and signifies explicit antisymmetrization of the indices
denoted under it.\footnote{
  Although $\sigma^{\mu\nu}$ is already antisymmetric,
  the sequence of covariant derivatives following is not,
  so this antisymmetrization following the denoted symmetrization is nontrivial.
  However, the additional antisymmetrizing terms this introduces into
  $\mathcal{O}_\sigma^{\nu\mu\mu_1\ldots\mu_s}$
  contract with $n_\mu n_{\mu_1} \ldots n_{\mu_s}$ to zero,
  so we are free to introduce them here.
  These terms are necessary for the local operator to transform under the
  $\left(\frac{n+2}{2},\frac{n}{2}\right)\oplus\left(\frac{n}{2},\frac{n+2}{2}\right)$
  representation of the Lorentz group.
  See Ref.~\cite{Geyer:1999uq} for details.
}
For the gluons, we have:
\begin{subequations}
  \begin{align}
    \mathcal{O}_{gV}^{\mu\nu\mu_1\ldots\mu_s}
    &=
    \underset{\{\mu\nu\mu_1\ldots\mu_s\}}{\mathcal{S}}
    2\mathrm{Tr}\left\{
      G^{\mu\pi}(0)
      \left(i\overleftrightarrow{\mathcal{D}}^{\mu_1}\right)
      \ldots
      \left(i\overleftrightarrow{\mathcal{D}}^{\mu_s}\right)
      G_{\pi}^{\phantom{\pi}\nu}(0)
      \right\} \,,
    \label{eqn:local:gluon:unpolarized}
    \\
    \mathcal{O}_{gA}^{\mu\nu\mu_1\ldots\mu_s}
    &=
    -i
    \underset{\{\mu\nu\mu_1\ldots\mu_s\}}{\mathcal{S}}
   2 \mathrm{Tr}\left\{
      G^{\mu\pi}(0)
      \left(i\overleftrightarrow{\mathcal{D}}^{\mu_1}\right)
      \ldots
      \left(i\overleftrightarrow{\mathcal{D}}^{\mu_s}\right)
      \widetilde{G}_{\pi}^{\phantom{\pi}\nu}(0)
      \right\} \,,
    \label{eqn:local:gluon:polarized}
    \\
    \mathcal{O}_{gT}^{\mu\nu\rho\sigma\mu_1\ldots\mu_s}
    &=
    \underset{[\mu\rho]}{\mathcal{A}}
    \underset{[\nu\sigma]}{\mathcal{A}}
    \underset{\{\rho\sigma\}}{\mathcal{S}}
    \underset{\{\mu\nu\mu_1\ldots\mu_s\}}{\mathcal{S}}
    2 \mathrm{Tr}\left\{
      G^{\mu\rho}(0)
      \left(i\overleftrightarrow{\mathcal{D}}^{\mu_1}\right)
      \ldots
      \left(i\overleftrightarrow{\mathcal{D}}^{\mu_s}\right)
      G^{\sigma\nu}(0)
      \right\} \,.
    \label{eqn:local:gluon:transversity}
  \end{align}
  \label{eqn:local:gluon}
\end{subequations}

These symmetrizations (and trace subtractions) serve two purposes:
first, they refine the local operators to transform under
irreducible representations of the Lorentz group,
allowing straightforward classification and decomposition of the operators;
and second, they ensure that none of the form factors in the
decomposition of matrix elements of the local operators contain
terms that will contract with $n_{\mu}n_{\mu_1}\ldots n_{\mu_s}$
to zero. This ensures that each of the generalized form factors
we find is actually present in the Mellin moment of a GPD.

The local operators defined in Eqs.~(\ref{eqn:local:quark}) and (\ref{eqn:local:gluon}),
like the bilocal operators we have derived them from,
have an additional dependence on the renormalization
scale $\mu^2$ which we have not notated.
Consequently, the form factors obtained from decomposing their matrix elements
will generally have dependence on the renormalization scale as well.
Electromagnetic form factors are a special exception to this rule.

\section{Counting generalized form factors}
\label{sec:gff} 
In this section,
we look at the decomposition of matrix elements of the local operators
given in Eqs.~(\ref{eqn:local:quark}) and (\ref{eqn:local:gluon})
when sandwiched between initial and final state kets
$|p,\lambda\rangle$ and $\langle p^\prime,\lambda^\prime|$, namely,
\begin{equation}
  \langle p^\prime,\lambda^\prime|\mathcal{O}^{\mu\nu\ldots}|p,\lambda\rangle
  \label{eqn:local:element}
\end{equation}
for a hadron $h$, where $\lambda (\lambda')$ is the light front helicity of the initial (final) state hadron.
These matrix elements can be decomposed into a number of Lorentz structures
containing the momenta $P$ and $\Delta$, and possibly Clifford matrices
sandwiched between spinors (in the spin-half case) or polarization vectors
(in the spin-one case). The Lorentz-invariant functions of the Lorentz scalar
$t = \Delta^2$ multiplying these structures
constitute the generalized form factors that we seek.

The number of GFFs that should appear in the decomposition of
Eq.~(\ref{eqn:local:element}) can be determined by using the method
outlined in Refs.~\cite{Ji:2000id,Hagler:2004yt}.
This method involves looking at the crossed channel
\begin{equation}
  \langle h\bar{h}|\mathcal{O}^{\mu\nu\ldots}|0\rangle
  \label{eqn:local:crossed}
\end{equation}
for hadron-antihadron production from the vacuum,
and determining the number of form factors for this process by matching the
$J^{PC}$ quantum numbers between the final state $\langle h\bar{h}|$
and the operator $\mathcal{O}^{\mu\nu\ldots}$.
More specifically, the procedure is as follows:
\begin{enumerate}
  \item For an operator $\mathcal{O}^{\mu\nu\ldots}$,
    the possible $J^{PC}$ quantum numbers are determined
    using its transformation properties under the Lorentz group.
  \item For each $J^{PC}$ quantum number, the total number of states
    ({\sl i.e.}, $L$ values) that are possible in an $h\bar{h}$ system
    with this $J^{PC}$ are counted.
  \item These counts for all possible $J^{PC}$ numbers are added up.
\end{enumerate}
The number of form factors obtained for the crossed channel is equal
to the number of GFFs for the original channel due to crossing symmetry.


\subsection{\texorpdfstring{$\boldsymbol{J^{PC}}$}{JPC} of local operators}
The $J^{PC}$ decompositions of the local operators of interest are independent
of the hadron. These decompositions have been derived elsewhere~\cite{Geyer:1999uq,Geyer:2000ig,Ji:2000id,Hagler:2004yt,Chen:2004cg,Pire:2014fwa},
and here we only recall the results.

The vector operators $O_{qV}^{\mu\mu_1\ldots\mu_s}$
and $O_{gV}^{\mu\nu\mu_1\ldots\mu_{s-1}}$ defined in
Eqs.~(\ref{eqn:local:quark:unpolarized}) and (\ref{eqn:local:gluon:unpolarized})
both transform under the $\left(\frac{s+1}{2},\frac{s+1}{2}\right)$
representation of the proper Lorentz group, and have $J^{PC}$ quantum numbers
\begin{align}
  J^{PC} = j^{ (-)^j (-)^{s+1} } \qquad : \qquad j \in \{0,1,\ldots,s+1\}\,.
\end{align}
Meanwhile, the axial vector operators
$O_{qA}^{\mu\mu_1\ldots\mu_s}$ and $O_{gA}^{\mu\nu\mu_1\ldots\mu_{s-1}}$
defined in Eqs.~(\ref{eqn:local:quark:polarized}) and (\ref{eqn:local:gluon:polarized})
also transform under the  $\left(\frac{s+1}{2},\frac{s+1}{2}\right)$
representation of the proper Lorentz group, but have opposite parity
and charge conjugation quantum numbers, giving
\begin{align}
  J^{PC} = j^{ (-)^{j+1} (-)^{s} } \qquad : \qquad j \in \{0,1,\ldots,s+1\} \,.
\end{align}
In both cases, operators associated with the $(s+1)$th Mellin moment of a quark GPD
and the $s$th moment of a gluon GPD transform the same way under the
Lorentz group and will accordingly have the same number of
generalized form factors.
As a consequence these operators mix under QCD evolution.
This will no longer be the case when we consider transversity operators.

The transversity operator for the quark, defined in
Eq.~(\ref{eqn:local:quark:transversity}), transforms under the
$\left(\frac{s+2}{2},\frac{s}{2}\right)\oplus\left(\frac{s}{2},\frac{s+2}{2}\right)$
representation of the proper Lorentz group. Both parity quantum numbers are
available for each $j$ value in the decomposition, giving two sequences of
allowed $J^{PC}$ numbers
\begin{align}
  J^{PC} = j^{ (-)^{j+1} (-)^{s+1} }
  , \qquad
  J^{PC} = j^{ (-)^j (-)^{s+1} }
  \qquad : \qquad j \in \{1,\ldots,s+1\} \,.
\end{align}

The helicity-flip operator for the gluon, defined in
Eq.~(\ref{eqn:local:gluon:transversity}), transforms under the
$\left(\frac{s+4}{2},\frac{s}{2}\right)\oplus\left(\frac{s}{2},\frac{s+4}{2}\right)$
representation of the proper Lorentz group.
As with the quark transversity, both parities contribute for each available $j$.
We thus again get two sequences of $J^{PC}$~\cite{Chen:2004cg}
\begin{align}
  J^{PC} = j^{ (-)^{j} (-)^{s} }
  , \qquad
  J^{PC} = j^{ (-)^{j+1} (-)^{s} }
  \qquad : \qquad j \in \{2,\ldots,s+2\}\,.
\end{align}


\subsection{\texorpdfstring{$\boldsymbol{J^{PC}}$}{JPC} counting and matching for spin-0}
As explained above,
the number of expected generalized form factors in the matrix element
of a local operator is counted by matching the number of $J^{PC}$ states
available to a hadron-antihadron state and the $J^{PC}$ decomposition 
of the local operator.
This matching scheme has been performed for spin-half hadrons extensively
elsewhere, so we will not repeat this for the spin-half case.
We will look in detail at the bosonic cases spin-0 and spin-1.

First, we consider spin-0 as a simple case.
The allowed $J^{PC}$ quantum numbers for a hadron-antihadron state
are determined by the relations for two-boson states,
\begin{align}
  &P=(-1)^L \,,
  &C=(-1)^{L+S} \,,
  \label{eqn:PC}
\end{align}
where $J=|L-S|,\ldots,L+S$. Since $S=0$ for a system of two spin-0
particles, one simply has $J=L$.
The allowed states are given by the sequence
$J^{PC} = j^{(-)^j(-)^j}$ ($j\geq0$)
with one state per $j$ value.

We now proceed to match the $J^{PC}$ sequence arising from
hadron-antihadron states with the $J^{PC}$ decomposition
of the local operators of interest.
We begin with the vector operators
$O_{qV}^{\mu\mu_1\ldots\mu_s}$ for quarks or
$O_{gV}^{\mu\nu\mu_1\ldots\mu_{s-1}}$ for gluons.
The available $J^{PC}$ states in the operator are $j^{(-)^j(-)^{s+1}}$,
with $j\in\{0,1,\ldots,s+1\}$.
The hadron-antihadron pair gives us states $J^{PC} = j^{(-)^j(-)^j}$,
with $j\geq0$.
Summing over all possible coincidences between these sequences, we get
\begin{align}
  N(s) = \sum_{j=0}^{s+1} \Theta\Big( (-1)^{s+j+1}=1 \Big)
  \,,
\end{align}
where $\Theta(P)=1$ if $P$ is true, and $\Theta(P)=0$ otherwise.
If we reindex the sum using $r=s+j+1$, we have
\begin{align}
  N(s)
  = \sum_{r=s+1}^{2s+2} \Theta\Big( (-1)^r=1 \Big)
  = \left\lfloor\frac{s+1}{2}\right\rfloor + 1 \,.
\end{align}
The number of generalized form factors for $s=0,1,2,3,\ldots$ should thus be
$1,2,2,3,\ldots$, which is exactly what is seen in existing literature
on pion GFFs (see, {\sl e.g.}, \cite{Broniowski:2008dy}).
Note that $s\geq0$ for quarks, and $s\geq1$ for gluons, since in the latter
case we are taking the $s$th Mellin moment.

We next consider the axial operators.
From their decomposition, we found
$J^{PC}=j^{(-)^{j+1}(-)^s}$.
There are no matches between this and the $J^{PC}$
of allowed hadron-antihadron states due to the mismatch in parity.
We thus reproduce the well-known result that the helicity-dependent
GPDs of spin-0 particles identically vanish.
The helicity-flip (transversity) GPDs, however, do not vanish for spin-0,
so we have two more cases---the quark and gluon transversities---to consider.

For quark transversity operators,
the cases $J^{PC} = j^{(-)^j(-)^{s+1}}$ and $j^{(-)^{j+1}(-)^{s+1}}$
must both be matched against $J^{PC} = j^{(-)^j(-)^j}$.
Clearly, only the first of these sequences has matches.
The limits are given by $j\in\{1,2,\ldots,s+1\}$.
Except for the lower limit in $j$ being different,
this looks identical to the helicity-independent case.
Summing over all possible matches gives us
\begin{align}
  N(s) = \sum_{j=1}^{s+1} \Theta\Big( (-1)^{s+j+1}=1 \Big)
  = \left\lfloor\frac{s}{2}\right\rfloor + 1
  \,,
\end{align}
producing the sequence 1, 1, 2, 2, \ldots for the number of GFFs
when taking the $(s+1)$th Mellin moment ($s\geq0$).
This sequence agrees with what is seen in Ref.~\cite{Dorokhov:2011ew}.

For gluon transversity operators, we consider
$\mathcal{O}_{gT}^{\mu\nu\alpha\beta\mu_1\ldots\mu_{s}}$,
which means (in contrast to the other gluon cases) we are taking the $(s+1)$th
Mellin moments of the transversity GPDs.
Doing so gives us the sequences
\begin{align}
  J^{PC} = j^{ (-)^{j} (-)^{s} }
  , \qquad
  J^{PC} = j^{ (-)^{j+1} (-)^{s} }
  \qquad : \qquad j \in \{2,\ldots,s+2\} \,.
\end{align}
Only the first of these sequences has matches with $J^{PC}=j^{(-)^j(-)^j}$.
The number of matches we get is
\begin{align}
  N(s) = \sum_{j=2}^{s+2} \Theta\Big( (-1)^{s+j}=1 \Big)
  = \left\lfloor\frac{s}{2}\right\rfloor + 1
  \,.
\end{align}
The number of GFFs for the $(s+1)$th moments of the gluon transversity GPDs
thus follows the sequence 1, 1, 2, 2, \ldots, which is the same as the number
of transversity GFFs for the quarks. However, unlike in the nonhelicity-flip
cases, the moments of the quark and gluon operators are not offset from
one another.
This curious fact was also noted for spin-half in Ref.~\cite{Chen:2004cg}.


\subsection{\texorpdfstring{$\boldsymbol{J^{PC}}$}{JPC} counting and matching for spin-1}
The allowed $J^{PC}$ for a hadron-antihadron state consisting of two
spin-1 particles are given by the relations (\ref{eqn:PC}) for two-boson states.
The limits for $J$ are given by $J=|L-S|,\ldots,L+S$,
and the possible values for $S$ are given by $S=0,1,2$.
This gives us three sequences of $J^{PC}$ quantum numbers---one for each $S$ value.
We can codify rules for counting the number of states as follows.

For the $S=0$ states, we replicate the spin-0 case.
We have $J=L$, giving us a sequence $J^{PC} = j^{(-)^j(-)^j}$ of allowed quantum numbers,
with one $L$ value for each $J^{PC}$.

For $S=1$, we have three sequences of states to count.
\begin{enumerate}
  \item States for which $j=L+1$, which begin at $L=0$.
    These have $J^{PC} = j^{(-)^{j+1}(-)^{j}}$ $(j\geq1)$.
  \item States for which $j=L$.
    These begin at $L=1$, since $j\geq|L-1|$ when $S=1$,
    forbidding us to get $j=0$ when $L=0$.
    These have $J^{PC} = j^{(-)^{j}(-)^{j+1}}$ $(j\geq1)$.
  \item States for which $j=|L-1|$.
    The $L=0$ state is already counted in the $j=L+1$ sequence,
    so we count only the $L\geq1$ states.
    These have $J^{PC} = j^{(-)^{j+1}(-)^{j}}$ $(j\geq0)$.
\end{enumerate}

For $S=2$, we have five sequences of states to count.
\begin{enumerate}
  \item States for which $j=L+2$, which begin at $L=0$.
    These have $J^{PC} = j^{(-)^{j}(-)^{j}}$ $(j\geq2)$.
  \item States for which $j=L+1$.
    These begin at $L=1$, since $j\geq|L-2|$ when $S=2$,
    forbidding us to get $j=1$ when $L=0$.
    These have $J^{PC} = j^{(-)^{j+1}(-)^{j+1}}$ $(j\geq2)$.
  \item States for which $j=L$, which begin at $L=1$
    (for reasons also relating to forbidden states).
    These have $J^{PC} = j^{(-)^{j}(-)^{j}}$ $(j\geq1)$.
  \item States for which $j=|L-1|$.
    The $L=0$ and $L=1$ states are both forbidden,
    so this sequence starts at $L=2$.
    These states have $J^{PC} = j^{(-)^{j+1}(-)^{j+1}}$ $(j\geq1)$.
  \item States for which $j=|L-2|$.
    The $L=0$ state is already counted in the $j=L+2$ sequence,
    and the $L=1$ state is already counted in the $j=L$ sequence,
    so we only count the $L\geq2$ states.
    These have $J^{PC} = j^{(-)^{j}(-)^{j}}$ $(j\geq0)$.
\end{enumerate}

Adding these counts together, we have the following number of $L$
values available in each $J^{PC}$ sequence:
\begin{subequations}
  \begin{align}
    & J^{PC} = j^{(-)^j(-)^j}         \qquad & : & \qquad 2 + \Theta(j\geq1) + \Theta(j\geq2) \,,
    \label{eqn:JPC:spin1:a} \\
    & J^{PC} = j^{(-)^{j+1}(-)^{j}}   \qquad & : & \qquad 1 + \Theta(j\geq1) \,,
    \label{eqn:JPC:spin1:b} \\
    & J^{PC} = j^{(-)^{j}(-)^{j+1}}   \qquad & : & \qquad \Theta(j\geq1) \,,
    \label{eqn:JPC:spin1:c} \\
    & J^{PC} = j^{(-)^{j+1}(-)^{j+1}} \qquad & : & \qquad \Theta(j\geq1) + \Theta(j\geq2) \,.
    \label{eqn:JPC:spin1:d}
  \end{align}
  \label{eqn:JPC:spin1}
\end{subequations}

\setlength{\tabcolsep}{8pt}

\begin{table}[ht] 
\caption{Possible $J^{PC}$ quantum numbers for $h\bar{h}$ states of spin-1 particles.}
\label{tab:jpc_spin1}
\begin{tabular}{ llllll } 
$S=0$\\
 \hline
$J^{PC}$& $0^{++}$ & $1^{--}$& $2^{++}$ & $3^{--}$& $4^{++}$\\
$L$ &0 &1 &2 &3 &4\\
 \hline
\end{tabular}

\vspace{0.5cm}
\begin{tabular}{ lllllllllll } 
$S=1$\\
 \hline
$J^{PC}$& $0^{-+}$ & $1^{+-}$& $1^{-+}$ & $2^{+-}$& $2^{-+}$ & $3^{+-}$& 
$3^{-+}$ & $4^{+-}$& $4^{-+}$&$\cdots$\\
$L$ &1 &0,2 &1 &2 &1,3 &2,4 &3 &4 &3,5&$\cdots$\\
 \hline
\end{tabular}

\vspace{0.5cm}
\begin{tabular}{ llllllllllll } 
$S=2$\\
 \hline
$J^{PC}$& $0^{++}$ & $1^{--}$& $1^{++}$ & $2^{--}$& $2^{++}$ & $3^{--}$& 
$3^{++}$ & $4^{--}$& $4^{++}$&$\cdots$\\
$L$ &2 &1,3 &2 &1,3 &0,2,4 &1,3,5 &2,4 &3,5 &2,4,6&$\cdots$\\
 \hline
\end{tabular}
\end{table}

As an illustrative guide, we tabulate in Table~\ref{tab:jpc_spin1} the allowed
$J^{PC}$ quantum numbers for all three sequences up to $J=4$.
The $L$ values are explicitly included, and one can confirm the formulas
given in Eqs.~(\ref{eqn:JPC:spin1}) for $J$ up to $J=4$ by counting the $L$ values
in this table.
When comparable, the $J^{PC}$ and $L$ values we find coincide with those
found in Refs.~\cite{Liuti:talk:2012,Liuti:2019srv}.

With the $J^{PC}$ sequences for spin-1 hadron-antihadron states in hand,
we now proceed to count matches between these and the $J^{PC}$ decompositions
of local operators.


\subsubsection{GFF counting for spin-1: Vector operators}

We first consider the number of GFFs arising from the $(s+1)$th moment
of the helicity-independent quark correlator,
or the $s$th moment of the gluon correlator.
The $J^{PC}$ decomposition of the relevant operator is
$J^{PC} = j^{(-)^{j}(-)^{s+1}}$, with $j\in\{0,1,\ldots,s+1\}$.
The $J^{PC}$ counts for the hadron-antihadron states are noted in
Eqs.~(\ref{eqn:JPC:spin1}).
Two of the sequences noted contribute, namely those from
Eqs.~(\ref{eqn:JPC:spin1:a}) and (\ref{eqn:JPC:spin1:c}).
Counting the number of matches we get gives
\begin{align}
  N(s) &=
  \sum_{j=1}^{s+1} 1
  + 2 \sum_{j=0}^{s+1} \Theta\Big( (-1)^{j+s+1}=1 \Big)
  + \sum_{j=2}^{s+1} \Theta\Big( (-1)^{j+s}=1 \Big)
  \notag \\ &=
  (s+1)
  + 2 \sum_{r=s+1}^{2s+2} \Theta\Big( (-1)^{r}=1 \Big)
  + \sum_{r=s+2}^{2s+1} \Theta\Big( (-1)^{r}=1 \Big)
  \notag \\ &=
  (s+1)
  + 2 \left( \left\lfloor\frac{s+1}{2}\right\rfloor + 1 \right)
  + \Theta(s\geq1) \left\lfloor\frac{s+1}{2}\right\rfloor
  \notag \\ &=
  3 + s + \Big( 2 + \Theta(s\geq1) \Big)\left\lfloor\frac{s+1}{2}\right\rfloor
  \notag \\ &=
  3\left( 1 + \left\lfloor\frac{s+1}{2}\right\rfloor \right) + s
  \,,
\end{align}
which produces the sequence 3, 7, 8, 12, 13, \ldots for the number of GFFs.
This gives us a pattern of numbers which alternatively increases by four and one.
For illustrative purposes, we provide in Table~\ref{tab:matching:vector}
an explicit tabulation of $J^{PC}$ matches for $s$ values up to $s=3$,
which require $J$ values up to $J=4$.

\begin{table}[ht]
  \caption{
  (Color online)
  Matches of $J^{PC}$ quantum numbers between the spin-1 hadron-antihadron
  spectrum and a local vector operator, for several values of $s$.
  The red subscript specifies the number of $L$ values associated with
  the $J^{PC}$ value tabulated. These subscripted numbers are summed
  to give the value in the rightmost column.
  }
  \label{tab:matching:vector}
\begin{tabular}{l|lllllll}
 \hline
 \diagbox{$s$}{$J$}
 & 0& 1 & 2 & 3 & 4 & $\cdots$ & No. of GFFs\\
 \hline
 0 & $0^{+-}$& $1^{--}_{\tddr{3}}$&&&&&3\\
 1 & $0^{++}_{\tddr{2}}$& $1^{-+}_{\tddr{1}}$ & $2^{++}_{\tddr{4}}$&&&&7\\
 2 & $0^{+-}$& $1^{--}_{\tddr{3}}$ & $2^{+-}_{\tddr{1}}$ & 
$3^{--}_{\tddr{4}}$&&&8\\
 3 & $0^{++}_{\tddr{2}}$& $1^{-+}_{\tddr{1}}$ & $2^{++}_{\tddr{4}}$ & 
$3^{-+}_{\tddr{1}}$ & $4^{++}_{\tddr{4}}$ &&12\\
 \ldots & \ldots & \ldots & \ldots & \ldots & \ldots & \ldots & \ldots \\
 \hline
\end{tabular}
\end{table}


\subsubsection{GFF counting for spin-1: Axial vector operators}

Next we consider the helicity-dependent correlators, namely the $(s+1)$th
moment of the quark correlator or the $s$th moment of the gluon correlator.
The $J^{PC}$ sequence for the operator is $j^{(-)^{j+1}(-)^{s}}$.
The hadron-antihadron states this can be matched with are those appearing
in Eqs.~(\ref{eqn:JPC:spin1:b}) and( \ref{eqn:JPC:spin1:d}).
Counting the number of matches we have gives us
\begin{align}
  N(s) &=
  \sum_{j=1}^{s+1} 1
  + \sum_{j=0}^{s+1} \Theta\Big( (-1)^{j+s}=1 \Big)
  + \sum_{j=2}^{s+1} \Theta\Big( (-1)^{j+s+1}=1 \Big)
  =
  2(s+1)
  \,.
\end{align}
An explicit tabulation for matches for $s$ up to $s=3$ is included in
Table~\ref{tab:matching:axial}.

\begin{table}[ht]
  \caption{
  Matches of $J^{PC}$ for the local axial vector operators.
  See caption of Table~\ref{tab:matching:vector} for more details.
  }
  \label{tab:matching:axial}
\begin{tabular}{l|lllllll}
 \hline
 \diagbox{$s$}{$J$}
 & 0& 1 & 2 & 3 & 4 & $\cdots$ & No. of GFFs\\
 \hline
 0 & $0^{-+}_{\tddr{1}}$& $1^{++}_{\tddr{1}}$&&&&&2\\
 1 & $0^{--}_{\tddr{0}}$& $1^{+-}_{\tddr{2}}$ & $2^{--}_{\tddr{2}}$&&&&4\\
 2 & $0^{-+}_{\tddr{1}}$& $1^{++}_{\tddr{1}}$ & $2^{-+}_{\tddr{2}}$ & 
$3^{++}_{\tddr{2}}$&&&6\\
 3 & $0^{--}_{\tddr{0}}$& $1^{+-}_{\tddr{2}}$ & $2^{--}_{\tddr{2}}$ & 
$3^{+-}_{\tddr{2}}$ & $4^{--}_{\tddr{2}}$ &&8\\
 \ldots & \ldots & \ldots & \ldots & \ldots & \ldots & \ldots & \ldots \\
 \hline
\end{tabular}
\end{table}


\subsubsection{GFF counting for spin-1: Quark transversity operators}

For the quark helicity-flip form factors, two sequences of $J^{PC}$ are
present in the decomposition of the operator:
$j^{(-)^{j}(-)^{s+1}}$ and $j^{(-)^{j+1}(-)^{s+1}}$.
Since both parties are present, all of the sequences for hadron-antihadron
states notated in Eqs.~(\ref{eqn:JPC:spin1}) contribute to the count.
Counting all of the matches gives us
\begin{align}
  N(s) &=
  \sum_{j=1}^{s+1} \left\{
    \Theta\Big( (-1)^{j+s+1}=1 \Big) \Big( 3 + 2\Theta(j\geq1) + \Theta(j\geq2) \Big)
    +  \Theta\Big( (-1)^{j+s}=1 \Big) \Big( 2\Theta(j\geq1) + \Theta(j\geq2) \Big)
    \right\}
  \notag \\ &=
  5 + 3\left( s + \left\lfloor\frac{s}{2}\right\rfloor \right)
  \,.
\end{align}
This follows the sequence 5, 8, 14, 17, 23, \ldots for the number of GFFs,
with the sequence alternatively increasing by three and six.
An explicit tabulation for matches for $s$ up to $s=3$ is included in
Table~\ref{tab:matching:transverse:quark}.

\begin{table}[ht]
  \caption{
  Matches of $J^{PC}$ for the local transverse quark operators.
  Two tabulations are present because there are two $J^{PC}$ sequences
  (with opposite parity) in the decomposition of the transversity operator.
  See caption of Table~\ref{tab:matching:vector} for more details.
  }
  \label{tab:matching:transverse:quark}
\begin{tabular}{l|llllll}
 \hline
 \diagbox{$s$}{$J$}
 & 1 & 2 & 3 & 4 & $\cdots$ & No,\\
 \hline
 0 & $1^{--}_{\tddr{3}}$&&&&&3\\
 1 & $1^{-+}_{\tddr{1}}$ & $2^{++}_{\tddr{4}}$&&&&5\\
 2 & $1^{--}_{\tddr{3}}$ & $2^{+-}_{\tddr{2}}$ & 
$3^{--}_{\tddr{4}}$&&&8\\
 3 & $1^{-+}_{\tddr{1}}$ & $2^{++}_{\tddr{4}}$ & 
$3^{-+}_{\tddr{1}}$ & $4^{++}_{\tddr{4}}$ &&10\\
 \ldots & \ldots & \ldots & \ldots & \ldots & \ldots & \ldots \\
 \hline
\end{tabular}
~~~
\begin{tabular}{l|llllll}
 \hline
 \diagbox{$s$}{$J$}
 & 1 & 2 & 3 & 4 & $\cdots$ & No.\\
 \hline
 0 & $1^{+-}_{\tddr{2}}$&&&&&2\\
 1 & $1^{++}_{\tddr{1}}$ & $2^{-+}_{\tddr{2}}$&&&&3\\
 2 & $1^{+-}_{\tddr{2}}$ & $2^{--}_{\tddr{2}}$ & 
$3^{+-}_{\tddr{2}}$&&&6\\
 3 & $1^{++}_{\tddr{1}}$ & $2^{-+}_{\tddr{2}}$ & 
$3^{++}_{\tddr{2}}$ & $4^{-+}_{\tddr{2}}$ &&7\\
 \ldots & \ldots & \ldots & \ldots & \ldots & \ldots & \ldots \\
 \hline
\end{tabular}
\end{table}


\subsubsection{GFF counting for spin-1: Gluon transversity operators}

Lastly, we look at helicity-flip correlators for the gluon.
As in the spin-0 case, we consider the $(s+1)$th Mellin moment,
not offsetting this like we did with the nonflip moments.
The $J^{PC}$ decomposition of the operators give us two sequences,
namely $j^{(-)^{j}(-)^{s}}$ and $j^{(-)^{j+1}(-)^{s}}$, with
$j\in\{2,3,\ldots,s+2\}$.
Since both parities are present in this decomposition,
all of the sequences in the hadron-antihadron state spectrum notated in
Eqs.~(\ref{eqn:JPC:spin1}) contribute to our counting.
Counting the matches, we find
\begin{align}
  N(s) &=
  \sum_{j=2}^{s+2} \left\{
    \Theta\Big( (-1)^{j+s}=1 \Big) \Big( 3 + 2\Theta(j\geq1) + \Theta(j\geq2) \Big)
    +  \Theta\Big( (-1)^{j+s+1}=1 \Big) \Big( 2\Theta(j\geq1) + \Theta(j\geq2) \Big)
    \right\}
  \notag \\ &=
  6 + 3\left( s + \left\lfloor\frac{s}{2}\right\rfloor \right)
  \,.
\end{align}
This produces the sequence 6, 9, 15, 18, 24, \ldots for the number of
gluon transversity GFFs.
Like with the quark transversity GFFs, this sequences alternatively
increases by three and six.
However, in contrast to the spin-0 and spin-half cases, the number of
quark and gluon transversity GFFs do not coincide.
Instead, for each value of $s$, there is one more gluon transversity GFF.
An explicit tabulation for matches for $s$ up to $s=3$ is included in
Table~\ref{tab:matching:transverse:gluon}.

\begin{table}[ht]
  \caption{
  Matches of $J^{PC}$ for the local transverse gluon operators.
  Two tabulations are present because there are two $J^{PC}$ sequences
  (with opposite parity) in the decomposition of the transversity operator.
  See caption of Table~\ref{tab:matching:vector} for more details.
  }
  \label{tab:matching:transverse:gluon}
\begin{tabular}{l|llllll}
 \hline
 \diagbox{$s$}{$J$}
 & 2 & 3 & 4 & 5 & $\cdots$ & No.\\
 \hline
 0 & $2^{++}_{\tddr{4}}$&&&&&4\\
 1 & $2^{+-}_{\tddr{1}}$ & $3^{--}_{\tddr{4}}$&&&&5\\
 2 & $2^{++}_{\tddr{4}}$ & $3^{-+}_{\tddr{1}}$ & 
$4^{++}_{\tddr{4}}$&&&9\\
 3 & $2^{+-}_{\tddr{1}}$ & $3^{--}_{\tddr{4}}$ & 
$4^{+-}_{\tddr{1}}$ & $5^{--}_{\tddr{4}}$ &&10\\
 \ldots & \ldots & \ldots & \ldots & \ldots & \ldots & \ldots \\
 \hline
\end{tabular}
~~~
\begin{tabular}{l|llllll}
 \hline
 \diagbox{$s$}{$J$}
 & 2 & 3 & 4 & 5 & $\cdots$ & No.\\
 \hline
 0 & $2^{-+}_{\tddr{2}}$&&&&&2\\
 1 & $2^{--}_{\tddr{2}}$ & $3^{+-}_{\tddr{2}}$&&&&4\\
 2 & $2^{-+}_{\tddr{2}}$ & $3^{++}_{\tddr{2}}$ & 
$4^{-+}_{\tddr{2}}$&&&6\\
 3 & $2^{--}_{\tddr{2}}$ & $3^{+-}_{\tddr{2}}$ & 
$4^{--}_{\tddr{2}}$ & $5^{+-}_{\tddr{2}}$ &&8\\
 \ldots & \ldots & \ldots & \ldots & \ldots & \ldots & \ldots \\
 \hline
\end{tabular}
\end{table}


\section{Results: generalized form factors and polynomiality}
\label{sec:polynomial}
In this section, we give explicit expressions for the matrix elements of the
local operators that appear in Mellin moments of light cone correlators
when the operators are sandwiched between kets for spin-1 hadrons.
The spin-0 and spin-half cases have been considered elsewhere.
We start by considering the operators with free indices
and then contract their decompositions with the appropriate number of $n$ vectors
and compare to the correlator decompositions
to obtain polynomiality sum rules for the GPDs.
The correspondence between our decompositions for local gluon currents and a second decomposition in the literature~\cite{Detmold:2017oqb} are discussed in Appendix~\ref{sec:comp}.

\subsection{Vector operators}

For the vector operator towers, we can write the following decomposition
\begin{multline}
  \langle p^\prime | \mathcal{O}_{qV}^{\mu\mu_1\ldots\mu_s} | p \rangle
  =
  \underset{\{\mu\mu_1\ldots\mu_s\}}{\mathcal{S}} \Bigg\{
  -(\epsilon \epsilon'^{*}) P^\mu
  \sum_{\substack{i=0\\\text{even}}}^s 
  \Delta^{\mu_1}\ldots \Delta^{\mu_i} P^{\mu_{i+1}}\ldots P^{\mu_s} 
  A^q_{s+1,i}(t)
  \\
  + \left[ \epsilon^{\mu}(\epsilon^{\prime*}P)+\epsilon^{\prime*\mu}(\epsilon  P)\right]
  \sum_{\substack{i=0\\\text{even}}}^s
  \Delta^{\mu_1}\ldots \Delta^{\mu_i} P^{\mu_{i+1}}\ldots  P^{\mu_s}
  B^q_{s+1,i}(t)
  \\
  - \frac{2(\epsilon'^{*}P)(\epsilon P)}{M^2} P^{\mu}
  \sum_{\substack{i=0\\\text{even}}}^s 
  \Delta^{\mu_1}\ldots \Delta^{\mu_i} P^{\mu_{i+1}}\ldots P^{\mu_s} 
  C^q_{s+1,i}(t)
  \\
  + \left[ \epsilon^{\mu}(\epsilon^{\prime*}P)-\epsilon^{\prime*\mu}(\epsilon  P)\right] \sum_{\substack{i=1\\\text{odd}}}^s
  \Delta^{\mu_1}\ldots \Delta^{\mu_i} P^{\mu_{i+1}}\ldots  P^{\mu_s}
  D^q_{s+1,i}(t)
  \\
  + M^2 \epsilon^{\mu} \epsilon'^{*\mu_1}
  \sum_{\substack{i=0\\\text{even}}}^{s-1}
  \Delta^{\mu_2}\ldots \Delta^{\mu_{i+1}} P^{\mu_{i+2}}\ldots P^{\mu_s} 
  E^q_{s+1,i}(t)
  \\
  -\text{mod}(s,2) \left[
    (\epsilon'^{*}\epsilon) F^q_{s+1}(t)
    + \frac{2(\epsilon'^{*}P)(\epsilon P)}{M^2} G^q_{s+1}(t)
    \right]
  \Delta^{\mu}\Delta^{\mu_1}\ldots \Delta^{\mu_s}
  \Bigg\}
  \label{eqn:decompose:vector}
  \,,
\end{multline}
where the number of factors $\Delta$ is related to the T-even or odd nature of 
the accompanying tensor,
and the last term contains two terms comparable to the $D$-term for the spin-1/2 case~\cite{Polyakov:1999gs}.  Polarization four-vectors of the initial (final) hadron are denoted by $\epsilon$ ($\epsilon'^{*}$), where the helicity index $\lambda$ ($\lambda'$) is implicit in both these vectors and the bra and ket. 
 
The number of GFFs appearing at each value of $s$ in Eq.~(\ref{eqn:decompose:vector})
can be counted as follows: each odd value of $s$, one gets an additional
$D_{s+1,i}$, $E_{s+1,i}$, $F_{s+1}$, and $G_{s+1}$, so the count increases by four.
At each even value, an additional
$A_{s+1,i}$, $B_{s+1,i}$, and $C_{s+1,i}$ appear,
but the $D$-term-like GFFs $F_{s+1}$ and $G_{s+1}$ drop out, so the count increases by one.
Only three GFFs are non-zero at $s=0$, namely, $A_{1,0}$, $B_{1,0}$, and $C_{1,0}$,
so the count starts at three, the sequence goes as 3, 7, 8, 12, 13, \ldots,
and can be written $s + 3\left(1+\lfloor\frac{s+1}{2}\rfloor\right)$.
This agrees with the number derived through $J^{PC}$ matching.

Combining the decomposition of Eq.~(\ref{eqn:decompose:vector}) with
that of Eq.~(\ref{eqn:gpd:vector})
gives the following sum rules for moments of quark GPDs:
\begin{subequations}
  \begin{align}
    & H^q_{1,s+1}(\xi,t)
    \equiv
    \int_{-1}^1 \mathrm{d}x \, x^s H_1(x,\xi,t)
    = \sum_{\substack{i=0\\\text{even}}}^s (-2\xi)^i A^q_{s+1,i}(t)
    + \frac{1}{3} \sum_{\substack{i=0\\\text{even}}}^{s-1} (-2\xi)^i E^q_{s+1,i}(t)
    + \text{mod}(s,2)(-2\xi)^{s+1} F^q_{n+1}(t) \,,
    \\
    & H^q_{2,s+1}(\xi,t)
    = \sum_{\substack{i=0\\\text{even}}}^s (-2\xi)^i B^q_{s+1,i}(t) \,,
    \\
    & H^q_{3,s+1}(\xi,t)
    = \sum_{\substack{i=0\\\text{even}}}^s (-2\xi)^i C^q_{s+1,i}(t)
    + \text{mod}(s,2)(-2\xi)^{s+1}G^q_{s+1}(t) \,,
    \\
    & H^q_{4,s+1}(\xi,t)
    = \sum_{\substack{i=1\\\text{odd}}}^s (-2\xi)^i D^q_{s+1,i}(t) \,,
    \\
    & H^q_{5,s+1}(\xi,t)
    = \sum_{\substack{i=0\\\text{even}}}^{s-1} (-2\xi)^i E^q_{s+1,i}(t) \,.
  \end{align}
  \label{eqn:polynomial:vector}
\end{subequations}

Note that $H_{5,1}=0$, which is related to the Close-Kumano sum rule~\cite{Close:1990zw}.
The electromagnetic structure function $b_1(x)$ appearing in the sum rule,
which states that $\int_0^1 b_1(x) \,\mathrm{d}x = 0$,
is related to the GPD $H_5$ at leading order and leading twist by
\begin{align}
  b_1(x,Q^2) =
  \sum_{q} e_q^2 \left[
    H_5^q(x,0,0;\mu^2=Q^2) - H_5^q(-x,0,0;\mu^2=Q^2)
    \right]
  ,
\end{align}
where $x\in[0,1]$ is the support region for $b_1$.
The Close-Kumano sum rule follows from $H^q_{5,1}=0$
if $H^q_5$ vanishes in the forward limit at negative $x$ values.
This sum rule can be violated if the sea carries tensor polarization~\cite{Close:1990zw},
but $H^q_{5,1}=0$ is an inviolate consequence of Lorentz symmetry.

The tower of local operators arising from the vector gluon correlator is related
to the tower from the quark operator we just explored,
but with the value of $s$ offset.
Specifically, the $(s+1)$th moment of the vector quark correlator has the same
Lorentz transformation properties and quantum numbers as the $s$th moment
of the vector gluon correlator, where $s\geq1$ in this context
since there is not a zeroth Mellin moment.
This tower thus has a familiar decomposition,
\begin{multline}
  \langle p^\prime | \mathcal{O}_{gV}^{\mu\mu_s\mu_1\ldots\mu_{s-1}}
  | p \rangle
  =
  \underset{\{\mu\mu_1\ldots\mu_s\}}{\mathcal{S}}
  2 
  \Bigg\{
  -(\epsilon \epsilon'^{*}) P^\mu
  \sum_{\substack{i=0\\\text{even}}}^s 
  \Delta^{\mu_1}\ldots \Delta^{\mu_i} P^{\mu_{i+1}}\ldots P^{\mu_s} 
  A^g_{s+1,i}(t)
  \\
  + \left[ \epsilon^{\mu}(\epsilon^{\prime*}P)+\epsilon^{\prime*\mu}(\epsilon  P)\right]
  \sum_{\substack{i=0\\\text{even}}}^s
  \Delta^{\mu_1}\ldots \Delta^{\mu_i} P^{\mu_{i+1}}\ldots  P^{\mu_s}
  B^g_{s+1,i}(t)
  \\
  - \frac{2(\epsilon'^{*}P)(\epsilon P)}{M^2} P^{\mu}
  \sum_{\substack{i=0\\\text{even}}}^s 
  \Delta^{\mu_1}\ldots \Delta^{\mu_i} P^{\mu_{i+1}}\ldots P^{\mu_s} 
  C^g_{s+1,i}(t)
  \\
  + \left[ \epsilon^{\mu}(\epsilon^{\prime*}P)-\epsilon^{\prime*\mu}(\epsilon  P)\right] \sum_{\substack{i=1\\\text{odd}}}^s
  \Delta^{\mu_1}\ldots \Delta^{\mu_i} P^{\mu_{i+1}}\ldots  P^{\mu_s}
  D^g_{s+1,i}(t)
  \\
  + M^2 \epsilon^{\mu} \epsilon'^{*\mu_1}
  \sum_{\substack{i=0\\\text{even}}}^{s-1}
  \Delta^{\mu_2}\ldots \Delta^{\mu_{i+1}} P^{\mu_{i+2}}\ldots P^{\mu_s} 
  E^g_{s+1,i}(t)
  \\
  -\text{mod}(s,2) \left[
    (\epsilon'^{*}\epsilon) F^g_{s+1}(t)
    + \frac{2(\epsilon'^{*}P)(\epsilon P)}{M^2} G^g_{s+1}(t)
    \right]
  \Delta^{\mu}\Delta^{\mu_1}\ldots \Delta^{\mu_s}
  \Bigg\}
  \label{eqn:decompose:vector:gluon}
  \,,
\end{multline}
where we have an extra factor of 2 relative to the quark case, by convention,
and where we have chosen to label the second index as $\mu_s$ rather than $\nu$
in order to make the correspondence with the quark decomposition clearer.
If we compare this to the definitions of the helicity-independent gluon GPDs,
we get the following polynomiality relations for odd values of $s$:
\begin{subequations}
  \begin{align}
    & H^g_{1,s+1}(\xi,t)
    \equiv
    \int_0^1 \mathrm{d}x \, x^{s-1} H_1(x,\xi,t)
    = \sum_{\substack{i=0\\\text{even}}}^s (-2\xi)^i A^g_{s+1,i}(t)
    + \frac{1}{3} \sum_{\substack{i=0\\\text{even}}}^{s-1} (-2\xi)^i E^g_{s+1,i}(t)
    + \text{mod}(s,2)(-2\xi)^{s+1} F^g_{n+1}(t) \,,
    \\
    & H^g_{2,s+1}(\xi,t)
    = \sum_{\substack{i=0\\\text{even}}}^s (-2\xi)^i B^g_{s+1,i}(t) \,,
    \\
    & H^g_{3,s+1}(\xi,t)
    = \sum_{\substack{i=0\\\text{even}}}^s (-2\xi)^i C^g_{s+1,i}(t)
    + \text{mod}(s,2)(-2\xi)^{s+1}G^g_{s+1}(t) \,, 
    \\
    & H^g_{4,s+1}(\xi,t)
    = \sum_{\substack{i=1\\\text{odd}}}^s (-2\xi)^i D^g_{s+1,i}(t) \,,
    \\
    & H^g_{5,s+1}(\xi,t)
    = \sum_{\substack{i=0\\\text{even}}}^{s-1} (-2\xi)^i E^g_{s+1,i}(t)
    \,,
  \end{align}
  \label{eqn:polynomial:vector:gluon}
\end{subequations}
where the reflection symmetry around $x=0$ of gluon GPDs was used to reduce the integration range
to $[0,1]$, and where the GPD moments for even $s$ are zero
because of this same symmetry.


\subsection{Axial vector operators}

For the axial vector operator towers, we can write the following decomposition:
\begin{multline}
  \langle p^\prime | \mathcal{O}_{qA}^{\mu\mu_1\ldots\mu_s} | p \rangle
  =
  \underset{\{\mu\mu_1\ldots\mu_s\}}{\mathcal{S}} \Bigg\{
  -i \epsilon^{\mu\nu\rho\sigma}
  \epsilon'^{*}_\nu \epsilon_\rho P_\sigma 
  \sum_{\substack{i=0\\\text{even}}}^s 
  \Delta^{\mu_1}\ldots \Delta^{\mu_i} P^{\mu_{i+1}}\ldots P^{\mu_s} 
  \widetilde{A}^q_{s+1,i}(t)
  \\
  + i \epsilon^{\mu\nu\rho\sigma} 
  \Delta_\nu P_\rho
  2\left[
    \frac{\epsilon_{\sigma}(\epsilon'^{*} P)
    + \epsilon'^{*}_{\sigma} (\epsilon  P)}{M^2}
    \right]
  \sum_{\substack{i=0\\\text{even}}}^s
  \Delta^{\mu_1}\ldots \Delta^{\mu_i} P^{\mu_{i+1}}\ldots P^{\mu_s}
  \widetilde{B}^q_{s+1,i}(t)
  \\
  + i \epsilon^{\mu\nu\rho\sigma} 
  \Delta_\nu P_\rho
  2\left[
    \frac{\epsilon_{\sigma}(\epsilon'^{*} P)
    - \epsilon'^{*}_{\sigma} (\epsilon P)}{M^2}
    \right]
  \sum_{\substack{i=1\\\text{odd}}}^s
  \Delta^{\mu_1}\ldots \Delta^{\mu_i} P^{\mu_{i+1}}\ldots P^{\mu_s}
  \widetilde{C}^q_{s+1,i}(t)
  \\
  + i \epsilon^{\mu\nu\rho\sigma} 
  \Delta_\nu P_\rho \left[\frac{
  \epsilon_{\sigma}\epsilon'^{*\mu_1} 
  +\epsilon'^{*}_{\sigma} \epsilon^{\mu_1}}{2}
  \right]\sum_{\substack{i=0\\\text{even}}}^{s-1}
  \Delta^{\mu_2}\ldots \Delta^{\mu_{i+1}} P^{\mu_{i+2}}\ldots P^{\mu_s} 
  \widetilde{D}^q_{s+1,i}(t)
  \Bigg\}
  \label{eqn:decompose:axial}
  \,,
\end{multline}
where the number of factors $\Delta$ is related to the T-even or odd nature of 
the accompanying remaining tensor.
The number of GFFs for any value of $s$ is $2(s+1)$,
which is equal to the number derived through $J^{PC}$ matching.

Combining the decomposition of Eq.~(\ref{eqn:decompose:axial}) with
that of Eq.~(\ref{eqn:gpd:axial})
gives the following sum rules for moments of helicity-dependent quark GPDs:
\begin{subequations}
  \begin{align}
    & \widetilde{H}^q_{1,s+1}(\xi,t)
    = \sum_{\substack{i=0\\\text{even}}}^s (-2\xi)^i \widetilde{A}^q_{s+1,i}(t)\,,
    \\
    & \widetilde{H}^q_{2,s+1}(\xi,t)
    = \sum_{\substack{i=0\\\text{even}}}^s (-2\xi)^i  \widetilde{B}^q_{s+1,i}(t) \,,
    \\
    & \widetilde{H}^q_{3,s+1}(\xi,t)
    = \sum_{\substack{i=1\\\text{odd}}}^s (-2\xi)^i  \widetilde{C}^q_{s+1,i}(t) \,,
    \\
    & \widetilde{H}^q_{4,s+1}(\xi,t)
    = \sum_{\substack{i=0\\\text{even}}}^{s-1} (-2\xi)^i \widetilde{D}^q_{n+1,i}(t)
  \,,
  \end{align}
  \label{eqn:polynomial:axial}
\end{subequations}
and we see that $\widetilde{H}^q_{3,1}=0$ and $\widetilde{H}^q_{4,1}=0$.

As in the vector case, the tower of local axial vector gluon operators matches up
with the axial vector quark operators, with $s$ offset to $(s-1)$.
We obtain the following decomposition for this gluon tower:
\begin{multline}
  \langle p^\prime | \mathcal{O}_{gA}^{\mu\mu_s\mu_1\ldots\mu_{s-1}} | p \rangle
  =
  \underset{\{\mu\mu_1\ldots\mu_s\}}{\mathcal{S}}
  2 
  \Bigg\{
  -i \epsilon^{\mu\nu\rho\sigma}
  \epsilon'^{*}_\nu \epsilon_\rho P_\sigma 
  \sum_{\substack{i=0\\\text{even}}}^s 
  \Delta^{\mu_1}\ldots \Delta^{\mu_i} P^{\mu_{i+1}}\ldots P^{\mu_s} 
  \widetilde{A}^g_{s+1,i}(t)
  \\
  + i \epsilon^{\mu\nu\rho\sigma} 
  \Delta_\nu P_\rho
  2\left[
    \frac{\epsilon_{\sigma}(\epsilon'^{*} P)
    + \epsilon'^{*}_{\sigma} (\epsilon  P)}{M^2}
    \right]
  \sum_{\substack{i=0\\\text{even}}}^s
  \Delta^{\mu_1}\ldots \Delta^{\mu_i} P^{\mu_{i+1}}\ldots P^{\mu_s}
  \widetilde{B}^g_{s+1,i}(t)
  \\
  + i \epsilon^{\mu\nu\rho\sigma} 
  \Delta_\nu P_\rho
  2\left[
    \frac{\epsilon_{\sigma}(\epsilon'^{*} P)
    - \epsilon'^{*}_{\sigma} (\epsilon P)}{M^2}
    \right]
  \sum_{\substack{i=1\\\text{odd}}}^s
  \Delta^{\mu_1}\ldots \Delta^{\mu_i} P^{\mu_{i+1}}\ldots P^{\mu_s}
  \widetilde{C}^g_{s+1,i}(t)
  \\
  + i \epsilon^{\mu\nu\rho\sigma} 
  \Delta_\nu P_\rho \left[\frac{
  \epsilon_{\sigma}\epsilon'^{*\mu_1} 
  +\epsilon'^{*}_{\sigma} \epsilon^{\mu_1}}{2}
  \right]\sum_{\substack{i=0\\\text{even}}}^{s-1}
  \Delta^{\mu_2}\ldots \Delta^{\mu_{i+1}} P^{\mu_{i+2}}\ldots P^{\mu_s} 
  \widetilde{D}^g_{s+1,i}(t)
  \Bigg\}
  \label{eqn:decompose:axial:gluon}
  \,,
\end{multline}
which leads us to the following sum rules for even values of $s$:
\begin{subequations}
  \begin{align}
    & \widetilde{H}^g_{1,s+1}(\xi,t)
    = \sum_{\substack{i=0\\\text{even}}}^s (-2\xi)^i \widetilde{A}^g_{s+1,i}(t) \,,
    \\
    & \widetilde{H}^g_{2,s+1}(\xi,t)
    = \sum_{\substack{i=0\\\text{even}}}^s (-2\xi)^i  \widetilde{B}^g_{s+1,i}(t) \,,
    \\
    & \widetilde{H}^g_{3,s+1}(\xi,t)
    = \sum_{\substack{i=1\\\text{odd}}}^s (-2\xi)^i  \widetilde{C}^g_{s+1,i}(t) \,,
    \\
    & \widetilde{H}^g_{4,s+1}(\xi,t)
    = \sum_{\substack{i=0\\\text{even}}}^{s-1} (-2\xi)^i \widetilde{D}^g_{n+1,i}(t)
  \,,
  \end{align}
  \label{eqn:polynomial:axial:gluon}
\end{subequations}
with the moments vanishing for odd $s$ since the helicity-dependent gluon
GPDs are odd under reflection about $x=0$.


\subsection{Quark transversity operators}

For the quark tensor operator towers, we can write the following decomposition:

\begin{multline}
  \langle p^\prime | \mathcal{O}_{qT}^{\mu\nu\mu_1\ldots\mu_s} | p \rangle
  =
  \underset{[\mu\nu]}{\mathcal{A}}
  \underset{\{\mu\mu_1\ldots\mu_s\}}{\mathcal{S}}
  2 
  \Bigg\{
  \frac{M}{2\sqrt{2}}
  \left(\epsilon'^{* \mu} \epsilon^{\nu } \right)
  \sum_{\substack{i=0\\\text{even}}}^s
  \Delta^{\mu_1}\ldots \Delta^{\mu_i} P^{\mu_{i+1}}\ldots P^{\mu_s} 
  A^{qT}_{s+1,i}(t)
  \\
  + \frac{M}{2\sqrt{2}}
  \left(\epsilon'^{*\mu} P^{\nu} \epsilon^{\mu_1}+ \epsilon^{\mu} P^{\nu} \epsilon'^{*\mu_1} \right)
  \sum_{\substack{i=1\\\text{odd}}}^{s-1}
  \Delta^{\mu_2}\ldots \Delta^{\mu_{i+1}} P^{\mu_{i+2}}\ldots P^{\mu_s} 
  B^{qT}_{s+1,i}(t)
  \\
  +\left[
    \frac{\epsilon'^{*\mu}\Delta^{\nu}(\epsilon P)}{M}
    - \frac{\epsilon^{\mu}\Delta^{\nu}(\epsilon'^{*} P)}{M}
    \right]
  \sum_{\substack{i=1\\\text{odd}}}^s
  \Delta^{\mu_1}\ldots \Delta^{\mu_{i}} P^{\mu_{i+1}}\ldots P^{\mu_s}
  C^{qT}_{s+1,i}(t)
  \\
  + \left[
    \frac{\epsilon'^{*\mu}\Delta^{\nu}(\epsilon P)}{M}
    + \frac{\epsilon^{\mu}\Delta^{\nu}(\epsilon'^{*} P)}{M}
    \right]
  \sum_{\substack{i=0\\\text{even}}}^s
  \Delta^{\mu_1}\ldots \Delta^{\mu_{i}} P^{\mu_{i+1}}\ldots P^{\mu_s}
  D^{qT}_{s+1,i}(t)
  \\
  + \frac{M}{2\sqrt{2}}
  \left[
    \epsilon'^{*\mu}\Delta^{\nu}\epsilon^{\mu_1}
    + \epsilon^{\mu}\Delta^{\nu}\epsilon'^{*\mu_1}
    \right]
  \sum_{\substack{i=0\\\text{even}}}^{s-1} 
  \Delta^{\mu_2}\ldots \Delta^{\mu_{i+1}} P^{\mu_{i+2}}\ldots P^{\mu_s}
  E^{qT}_{s+1,i}(t)
  \\
  + \frac{P^{\mu}\Delta^{\nu}}{M} \left(\epsilon'^{*} \epsilon\right) 
  \sum_{\substack{i=0\\\text{even}}}^s
  \Delta^{\mu_1}\ldots \Delta^{\mu_{i}} P^{\mu_{i+1}}\ldots P^{\mu_s}
  F^{qT}_{s+1,i}(t)
  \\
  + \frac{P^{\mu}\Delta^{\nu}}{M} \frac{(\epsilon'^{*}P)(\epsilon P)}{M^2} 
  \sum_{\substack{i=0\\\text{even}}}^s
  \Delta^{\mu_1}\ldots \Delta^{\mu_{i}} P^{\mu_{i+1}}\ldots P^{\mu_s}
  G^{qT}_{s+1,i}(t)
  \\
  + \left[
    \frac{\epsilon'^{*\mu}P^{\nu}(\epsilon P)}{M}
    + \frac{\epsilon^{\mu}P^{\nu}(\epsilon'^{*} P)}{M}
    \right]
  \sum_{\substack{i=1\\\text{odd}}}^s
  \Delta^{\mu_1}\ldots \Delta^{\mu_{i}} P^{\mu_{i+1}}\ldots P^{\mu_s}
  H^{qT}_{s+1,i}(t)
  \\
  + \left[
    \frac{\epsilon'^{*\mu}P^{\nu}(\epsilon P)}{M}
    - \frac{\epsilon^{\mu}P^{\nu}(\epsilon'^{*} P)}{M}
    \right]
    \sum_{\substack{i=0\\\text{even}}}^s
  \Delta^{\mu_1}\ldots \Delta^{\mu_{i}} P^{\mu_{i+1}}\ldots P^{\mu_s}
  I^{qT}_{n+1,i}(t)
  \Bigg\}
  \,.
  \label{eqn:decompose:transverse:quark}
\end{multline}
The number of GFFs present for any particular $s$ value is
$5 + 3\left(s + \lfloor\frac{s}{2}\rfloor\right)$,
which is exactly what was found above through $J^{PC}$ matching.
One can determine this number by noting that four of the Lorentz structures
in this decomposition are zero when $s=0$, that three GFFs are added with each
power of $s$, and that six are added with each even power of $s$---meaning that,
as observed in the $J^{PC}$ matching section, the number of GFFs follows a
sequence that starts with 5 and increases alternatively by three and six:
5, 8, 14, 17, 23, \ldots

Unlike with the vector and axial vector operator towers,
there is no special correspondence with gluon transversity GFFs.

Using Eq.~(\ref{eqn:decompose:transverse:quark}) with the definition
of the quark transversity GPDs in Eq.~(\ref{eqn:gpd:transversity:quark})
yields the following sum rules for Mellin moments of
quark transversity GPDs:
\begin{subequations}
  \begin{align}
    & H^{qT}_{1,s+1}(\xi,t) =
    \sum_{\substack{i=0\\\text{even}}}^s (-2\xi)^i A^{qT}_{s+1,i}(t) \,,
    \\
    & H^{qT}_{2,s+1}(\xi,t) =
    \sum_{\substack{i=1\\\text{odd}}}^{s-1} (-2\xi)^i B^{qT}_{s+1,i}(t) \,,
    \\
    & H^{qT}_{3,s+1}(\xi,t) =
    \sum_{\substack{i=1\\\text{odd}}}^{s} (-2\xi)^i C^{qT}_{s+1,i}(t) \,,
    \\
    & H^{qT}_{4,s+1}(\xi,t) =
    \sum_{\substack{i=0\\\text{even}}}^{s} (-2\xi)^i D^{qT}_{s+1,i}(t) \,,
    \\
    & H^{qT}_{5,s+1}(\xi,t) =
    \sum_{\substack{i=0\\\text{even}}}^{s-1} (-2\xi)^i E^{qT}_{s+1,i}(t) \,,
    \\
    & H^{qT}_{6,s+1}(\xi,t) =
    \sum_{\substack{i=0\\\text{even}}}^s (-2\xi)^i F^{qT}_{s+1,i}(t) \,,
    \\
    & H^{qT}_{7,s+1}(\xi,t) =
    \sum_{\substack{i=0\\\text{even}}}^s (-2\xi)^i G^{qT}_{s+1,i}(t) \,,
    \\
    & H^{qT}_{8,s+1}(\xi,t) =
    \sum_{\substack{i=1\\\text{odd}}}^s (-2\xi)^i H^{qT}_{s+1,i}(t) \,,
    \\
    & H^{qT}_{9,s+1}(\xi,t) =
    \sum_{\substack{i=0\\\text{even}}}^s (-2\xi)^i I^{qT}_{s+1,i} 
\,,
  \end{align}
\end{subequations}
where we note that
$H_{2,1}^{qT} = H_{3,1}^{qT} = H_{5,1}^{qT} = H_{8,1}^{qT} = 0$
are the four first Mellin moments that vanish.


\subsection{Gluon transversity operators}

We lastly look at the tower of gluon transversity operators,
for which we can write the following decomposition:
\begin{multline}
  \langle p^\prime | \mathcal{O}_{gT}^{\mu\nu\rho\sigma\mu_1\ldots\mu_s} | p \rangle
  =
  \underset{[\mu \rho ]}{\mathcal{A}}
  \underset{[ \sigma \nu ]}{\mathcal{A}}
  \underset{\{\rho \sigma \}}{\mathcal{S}}
  \underset{\{\mu \nu \mu_1\ldots \mu_{s} \}}{\mathcal{S}}
  8 
  \Bigg\{
  \frac{P^{\nu}\Delta^{\sigma}}{M} \Bigg[
  \frac{M}{2\sqrt{2}} \left(\epsilon'^{*\mu} \epsilon^{\rho} \right)
  \sum_{\substack{i=0\\\text{even}}}^s 
  \Delta^{\mu_1}\ldots \Delta^{\mu_i} P^{\mu_{i+1}}\ldots P^{\mu_s} 
  A^{gT}_{s+1,i}(t)
  \\
  + \frac{M}{2\sqrt{2}}
  \left(
    \epsilon'^{*\mu} P^{\rho} \epsilon^{\mu_1}
    + \epsilon^{\mu} P^{\rho} \epsilon'^{*\mu_1}
  \right)
  \sum_{\substack{i=1\\\text{odd}}}^{s-1}
  \Delta^{\mu_2}\ldots \Delta^{\mu_{i+1}} P^{\mu_{i+2}}\ldots P^{\mu_s}
  B^{gT}_{s+1,i}(t)
  \\
  + \left(
    \frac{\epsilon'^{*\mu}\Delta^{\rho}(\epsilon P)}{M}
    - \frac{\epsilon^{\mu}\Delta^{\rho}(\epsilon'^{*} P)}{M}
    \right)
  \sum_{\substack{i=1\\\text{odd}}}^s
  \Delta^{\mu_1}\ldots \Delta^{\mu_{i}} P^{\mu_{i+1}}\ldots P^{\mu_s}
  C^{gT}_{s+1,i}(t)
  \\
  + \left(
    \frac{\epsilon'^{*\mu}\Delta^{\rho}(\epsilon P)}{M}
    + \frac{\epsilon^{\mu}\Delta^{\rho}(\epsilon'^{*} P)}{M}
    \right)
  \sum_{\substack{i=0\\\text{even}}}^s
  \Delta^{\mu_1}\ldots \Delta^{\mu_{i}} P^{\mu_{i+1}}\ldots P^{\mu_s}
  D^{gT}_{s+1,i}(t)
  \\
  + \frac{P^{\mu}\Delta^{\rho}}{M} \frac{(\epsilon'^{*}P)(\epsilon P)}{M^2} 
  \sum_{\substack{i=0\\\text{even}}}^s
  \Delta^{\mu_1}\ldots \Delta^{\mu_{i}} P^{\mu_{i+1}}\ldots P^{\mu_s}
  G^{gT}_{s+1,i}(t)
  \\
  + \left(
    \frac{\epsilon'^{*\mu}P^{\rho}(\epsilon P)}{M}
    + \frac{\epsilon^{\mu}P^{\rho}(\epsilon'^{*} P)}{M}
    \right)
  \sum_{\substack{i=1\\\text{odd}}}^s
  \Delta^{\mu_1}\ldots \Delta^{\mu_{i}} P^{\mu_{i+1}}\ldots P^{\mu_s}
  H^{gT}_{s+1,i}(t)
  \\
  + \left(
    \frac{\epsilon'^{*\mu}P^{\rho}(\epsilon P)}{M}
    - \frac{\epsilon^{\mu}P^{\rho}(\epsilon'^{*} P)}{M}
    \right)
  \sum_{\substack{i=0\\\text{even}}}^s
  \Delta^{\mu_1}\ldots \Delta^{\mu_{i}} P^{\mu_{i+1}}\ldots P^{\mu_s}
  I^{gT}_{s+1,i}(t)
  \Bigg] 
  \\
  - \epsilon'^{*\mu}P^{\rho} \epsilon^{\nu}P^{\sigma}
  \sum_{\substack{i=0\\\text{even}}}^{s}
  \Delta^{\mu_1}\ldots \Delta^{\mu_{i}} P^{\mu_{i+1}}\ldots P^{\mu_s}
  E^{gT}_{s+1,i}(t)
  \\
  + \frac{1}{4} \epsilon'^{*\mu}\Delta^{\rho} \epsilon^{\nu}\Delta^{\sigma} 
  \sum_{\substack{i=0\\\text{even}}}^s
  \Delta^{\mu_1}\ldots \Delta^{\mu_{i}} P^{\mu_{i+1}}\ldots P^{\mu_s}
  F^{gT}_{s+1,i}(t) \Bigg\}
  \,.
  \label{eqn:decompose:transverse:gluon}
\end{multline}
One can observe that the number of GFFs arising from this decomposition
at order $(s+1)$ is $3 + 3\left(s + \lfloor\frac{s}{2}\rfloor\right)$
by noticing that there are six GFFs when $s=0$, and that the number of GFFs
increases by one at odd $s$ and six at even $s$.
This gives the sequence 6, 9, 15, 18, 24, \ldots,
which is exactly the number of GFFs derived from $J^{PC}$ matching above.

Finally, this gives us the following relations for
the moments of the gluon transversity GPDs,
with $s\geq 0$ and $s$ even:
\begin{subequations}
  \begin{align}
    & H^{gT}_{1,s+1}(\xi,t)
    \equiv \int_0^1 \mathrm{d}x \, x^{s} H^{gT}_1(x,\xi,t) =
    \sum_{\substack{i=0\\\text{even}}}^s (-2\xi)^i A^{gT}_{s+1,i}(t) \,,
    \\
    & H^{gT}_{2,s+1}(\xi,t) =
    \sum_{\substack{i=1\\\text{odd}}}^{s-1} (-2\xi)^i B^{gT}_{s+1,i}(t) \,,
    \\
    & H^{gT}_{3,s+1}(\xi,t) =
    \sum_{\substack{i=1\\\text{odd}}}^{s} (-2\xi)^i C^{gT}_{s+1,i}(t) \,,
    \\
    & H^{gT}_{4,s+1}(\xi,t) =
    \sum_{\substack{i=0\\\text{even}}}^{s} (-2\xi)^i D^{gT}_{s+1,i}(t) \,,
    \\
    & H^{gT}_{5,s+1}(\xi,t) =
    \sum_{\substack{i=0\\\text{even}}}^{s} (-2\xi)^i E^{gT}_{s+1,i}(t) \,,
    \\
    & H^{gT}_{6,s+1}(\xi,t) =
    \sum_{\substack{i=0\\\text{even}}}^s (-2\xi)^i F^{gT}_{s+1,i}(t) \,,
    \\
    & H^{gT}_{7,s+1}(\xi,t) =
    \sum_{\substack{i=0\\\text{even}}}^s (-2\xi)^i G^{gT}_{s+1,i}(t) \,,
    \\
    & H^{gT}_{8,s+1}(\xi,t) =
    \sum_{\substack{i=1\\\text{odd}}}^s (-2\xi)^i H^{gT}_{s+1,i}(t) \,,
    \\
    & H^{gT}_{9,s+1}(\xi,t) =
    \sum_{\substack{i=0\\\text{even}}}^s (-2\xi)^i I^{gT}_{s+1,i}(t)
    \,,
  \end{align}
\end{subequations}
where $H_{2,1}^{gT} = H_{3,1}^{gT} = H_{8,1}^{gT} = 0$ are the three first
Mellin moments that vanish.
Moments with odd $s$ are zero due to the reflection symmetry
of the gluon transversity GPDs about $x=0$.

\section{Conclusion}
\label{sec:concl}
In this work, we obtained polynomiality sum rules for spin-1 targets.
This was accomplished by decomposing off-forward matrix elements of the local
currents that appear in Mellin moments of bilocal operators.
Thus, as a byproduct of this derivation, we have also obtained the decomposition
of said local currents into independent generalized form factors.
The method of Ji and Lebed~\cite{Ji:2000id} was used to count the number of
independent generalized form factors that should appear in the decomposition
of each local current, and we find agreement with our results.

In principle, such work could be extended to systems with greater spin.
There exist spin-$3/2$ nuclei such as Lithium-7 which may be of future
experimental interest, such as in studies of the polarized EMC effect.
On an alternative route, the meaning of the generalized form factors appearing
in the Mellin moments of GPDs for spin-1 systems warrants more in-depth exploration.
The form factors appearing in the second Mellin moment of the helicity-independent GPDs
appear in the Lorentz-covariant decomposition of the energy-momentum tensor
and encode properties of great interest, such as the distribution of mass,
angular momentum, and forces (including shear and pressure forces) inside the hadron.

One curious feature of spin-1 targets,
which contrasts with spin-0 and spin-$1/2$ targets,
is the appearance of two independent ``$D$-like'' terms,
one each in the second Mellin moment of $H_1$ and $H_3$.
Two form factors may be necessary to describe the distribution of forces inside
a hadron with more complicated structure,
including three helicity states and a quadrupole moment.
This, and other aspects of the spin-1 energy-momentum tensor,
will be the focus of a future work.

\section*{ACKNOWLEDGMENTS}

A.F. and W.C. would like to thank the organizers of the ECT$^*$ workshop \emph{Exposing Novel Quark and Gluon Effects in Nuclei} (where discussions led to this collaboration), Simonetta Liuti for useful discussions during and after the workshop, and Misak Sargsian for
encouraging this work.
We thank W.\ Detmold, D.\ Pefkou and P.\ Shanahan for a careful reading of the manuscript and illuminating discussions.
A.F.\  was supported by the U.S.\ Department of Energy, Office of Science, Office of Nuclear Physics, contract no.\ DE-AC02-06CH11357, and an LDRD initiative at Argonne National Laboratory under Project No.\ 2017-058-N0.

\appendix
\section{CONVENTIONS USED}
\label{sec:conventions}

Since there are variations on conventions used in the GPD literature,
we feel it to be prudent to lay out the conventions used in this work
here.
The four-momenta $p$ and $p^\prime$ are carried by the initial and final
state hadron, respectively.
We additionally use 
\begin{align}
 &P = \frac{p+p^\prime}{2} \,, &\Delta = p^\prime-p\,,
\end{align}
for the average hadron momentum and momentum transfer, with Mandelstam variable $t=\Delta^2$.
We use null vectors $n$ and $\bar{n}$ to define the light cone,
with $(n \bar{n})=1$. These define the ``plus'' and ``minus'' components
of four-vectors as $x^+ = (x  n)$ and $x^- = (x \bar{n})$.
The skewness is defined as 
\begin{equation} \label{eq:xi}
 \xi = -\frac{(n\Delta)}{2(Pn)}\,.
\end{equation}

For the gluons, the dual field strength tensor can be obtained through
\begin{align}
  \widetilde{G}_{\mu\nu} = \frac{1}{2} \epsilon_{\mu\nu\rho\sigma} G^{\rho\sigma}
  ,
\end{align}
where we use the convention $\epsilon_{0123} = +1$.

As the order of symmetrization and antisymmetrization operations on the Lorentz tensors matters in this work, we will abstain from using the $[]$ and $\{\}$ notation within indices and use explicit operators
\begin{align}\label{eq:symm}
    &\underset{\{\mu_1\ldots\mu_s\}}{\mathcal{S}} T^{\mu_1 \ldots \mu_s}= \tfrac{1}{n!} \sum_{\sigma} T^{\sigma(\mu_1)\ldots \sigma(\mu_1)}\, - \,\text{removal of traces}\,,\\
    &\underset{[\mu_1\ldots\mu_s]}{\mathcal{A}} T^{\mu_1 \ldots \mu_s}= \tfrac{1}{n!} \sum_{\sigma} \text{sign}(\sigma) T^{\sigma(\mu_1)\ldots \sigma(\mu_1)} \qquad \text{for} \, s \leq 4\,,\label{eq:antisymm}
\end{align}
where $\sigma$ are permutations on the set $\{\mu_1,\ldots,\mu_n\}$, and the sign of $\sigma$ is determined by an even ($+1$) or odd ($-1$) number of permutations.
The ``removal of traces'' refers specifically to subtracting off all possible combinations of
pairwise contractions ({\sl i.e.}, traces) of indices in the symmetrized tensor;
because of a peculiarity of orthogonal groups (such as the Lorentz group), this is needed
for the symmetrized tensor to transform under an irreducible representation of the group
(see Chapter 10-5 of Ref.~\cite{Hammermesh} for details).

\section{COMPARISON WITH A SECOND GFF DECOMPOSITION}
\label{sec:comp}

For gluons and spin-1 targets, the GFF decomposition has also been written down for all leading-twist cases in Ref.~\cite{Detmold:2017oqb} [Eqs.~(7) and (8) and Appendix B therein, see also the Erratum], denoted [DPS] in the following.  Below, we summarize the correspondences between the form factors appearing in the two decompositions.
Note that Ref.~\cite{Detmold:2017oqb} does not provide a decomposition in the minimal linearly independent set of GFFs; hence the counting of form factors provided in Ref.~\cite{Detmold:2017oqb} differs from ours.


For the vector operator, we can make the following identification between Eq.~(7) [DPS] and our Eq.~(\ref{eqn:decompose:vector:gluon}):
\begin{align}
&B_{1,i}^{(s+1)}(t) = -2E^g_{s+1,i}(t)\,, 
&B_{2,i}^{(s+1)}(t) =  2A^g_{s+1,i}(t)\,, 
&&B_{3,s-1}^{(s+1)}(t) =2 F^g_{s+1}(t)\,, 
 &&B_{4,i}^{(s+1)}(t) = -2B^g_{s+1,i}(t)\,, \nonumber\\
&B_{5,i}^{(s+1)}(t) = -2D^g_{s+1,i}(t) \,,
&B_{6,i}^{(s+1)}(t) = 4C^g_{s+1,i}(t) \,,
&&B_{7,s-1}^{(s+1)}(t) = 4G^g_{s+1}(t) \,.
\end{align}
The [DPS] GFFs $B_{3,s-1}^{(s+1)},B_{7,s-1}^{(s+1)}$ correspond to our $D$-terms, and we can also identify $B_{7,s-2}^{(s+1)}=4C^g_{s+1,s}$ and $B_{3,s-2}^{(s+1)}=2A^g_{s+1,s} $. The other GFFs $(2\leq i \leq s-1)$ in these towers 3 and 7 are also present in the $B_{2,m}^{(s+1)}, B_{6,m}^{(s+1)}$ towers respectively:
\begin{align}
&B_{2,i}^{(s+1)}(t) = B_{3,i-2}^{(s+1)}(t)\,,
&B_{6,i}^{(s+1)}(t) = B_{7,i-2}^{(s+1)}(t) \,.
\end{align}


For the axial vector operators, we compare [DPS] Appendix B with our
Eq.~(\ref{eqn:decompose:axial:gluon}).
This results in the following direct correspondences:
\begin{align}
    &\widetilde{B}_{1,i}^{(s+1)}(t)=-2\widetilde{A}^g_{s+1,i}(t)\,,
    &\widetilde{B}_{3,i}^{(s+1)}(t)=-\widetilde{D}^g_{s+1,i}(t)\,,
\end{align}
The [DPS] towers $\widetilde{B}_{2,i}^{(s+1)}(t)$ and $\widetilde{B}_{4,i}^{(s+1)}(t)$ can be related to our GFFs after application of the Schouten identities.

For the tensor currents, the identification between [DPS] Eq.~(8) and our Eq.~(\ref{eqn:decompose:transverse:gluon}) is as follows:
\begin{align}
&A^{(s+2)}_{1,i} = 2 E^{gT}_{s+1,i}\,,
&A^{(s+2)}_{2,i} = -\tfrac{1}{2} F^{gT}_{s+1,i}\,,\nonumber\\
&A^{(s+2)}_{3,i} = \tfrac{1}{\sqrt{2}} A^{gT}_{s+1,i}\,,
&A^{(s+2)}_{4,i} = \tfrac{1}{\sqrt{2}} A^{gT}_{s+1,i}\,,\nonumber\\
&A^{(s+2)}_{5,i} =  -2D^{gT}_{s+1,i}\,,
&A^{(s+2)}_{6,i} =  -2I^{gT}_{s+1,i}\,,\nonumber\\
&A^{(s+2)}_{8,i} = 2G^{gT}_{s+1,i}\,,
&A^{(s+2)}_{9,i} = \tfrac{1}{\sqrt{2}}B^{gT}_{s+1,i}\,,\nonumber\\
&A^{(s+2)}_{10,i} =  2C^{gT}_{s+1,i}\,,
&A^{(s+2)}_{11,i} =  2H^{gT}_{s+1,i}\,,\nonumber\\
\end{align}
\begin{multline}
    - A^{(s+2)}_{7,i} + \frac{A^{(s+2)}_{7,i-2}}{4} = 2D^{gT}_{s+1,i}-\frac{D^{gT}_{s+1,i-2}}{2} -4I^{gT}_{s+1,i}+ I^{gT}_{s+1,i-2}-2\frac{t}{M^2}E^{gT}_{s+1,i}-2\left(1-\frac{t}{4M^2}\right) E^{gT}_{s+1,i-2}\\ \qquad -2\left(4+\frac{t}{M^2}\right)F^{gT}_{s+1,i}+\frac{t}{2M^2}F^{gT}_{s+1,i-2}\,,
\end{multline}
Of the towers [DPS] list, two are not linearly independent of the other nine $\left(A^{(s+2)}_{7,i}~\text{and}~A^{(s+2)}_{4,i}\right)$.  

\bibliography{main.bib}

\end{document}